\documentclass[aps,prb,twocolumn,superscriptaddress,floatfix,a4]{revtex4}
\usepackage{epsfig,amsmath,amssymb,color}
\begin{document}

\title{Field-dependent spin and heat conductivities of dimerized spin-$1/2$ chains}
\author{S. Langer}
\affiliation{Department of Physics and Arnold Sommerfeld Center for Theoretical Physics, Ludwig-Maximilians-Universit\"at M\"unchen, D-80333 M\"unchen, Germany}
\author{R. Darradi} 
\affiliation{Institut f\"ur Theoretische Physik, Technische Universit\"at Braunschweig, D-38106 Braunschweig, Germany}
\author{F. Heidrich-Meisner} 
\affiliation{Department of Physics and Arnold Sommerfeld Center for Theoretical Physics, Ludwig-Maximilians-Universit\"at M\"unchen, D-80333 M\"unchen, Germany}
\author{W. Brenig} 
\affiliation{Institut f\"ur Theoretische Physik, Technische Universit\"at Braunschweig, D-38106 Braunschweig, Germany}

\date{\today}

\begin{abstract} We study the spin and heat conductivity of dimerized spin-1/2
chains in homogeneous magnetic fields at finite temperatures. At zero
temperature, the model undergoes two field-induced quantum phase transitions
from a dimerized, into a Luttinger, and finally into a fully polarized
phase. We search for signatures of these transitions in the spin and heat
conductivities. Using exact diagonalization, we calculate the Drude weights, the
frequency dependence of the conductivities, and the corresponding integrated
spectral weights. As a main result, we demonstrate that both the spin and heat
conductivity are enhanced in the gapless phase and most notably at low
frequencies. In the case of the thermal conductivity, however, the
field-induced increase seen in the bare transport coefficients is suppressed by
magnetothermal effects, caused by the coupling of the heat and spin current in
finite magnetic fields. Our results complement recent magnetic
transport experiments on spin ladder materials with sufficiently small exchange
couplings allowing access to the field-induced transitions. \end{abstract}

\maketitle
\section{Introduction}
While low-dimensional quantum magnets have been intensely studied for the past
decade, the transport properties still pose viable challenges for experimental
and theoretical physicists. \cite{zotos-review,hm07a,hess07a,sologubenko07a} On
the theoretical side, the ground-state properties of low-dimensional spin
systems are very well studied by means of powerful techniques such as the
Bethe ansatz, bosonization, the density matrix renormalization group (DMRG) method, quantum Monte Carlo (QMC), and exact diagonalization
(see Refs.~\onlinecite{lnp645,everts03,dmrgmain} for reviews). Finite-temperature transport
properties, though, remain an exciting and active field of research. One of the
best established results is the ballistic thermal transport in the
integrable $XXZ$ spin-1/2 chain.\cite{zotos97,kluemper02,sakai03,hm03} In general,
however, transport properties are not that easily obtained, especially far from
equilibrium or in non-integrable models, but even spin transport in exactly
solvable models is still an active field of research with open pending
questions.\cite{zotos99,hm03,benz05,sirker09,grossjohann10,fujimoto03,andrei98,steinigeweg09}
In the non-equilibrium case, one has to resort to numerical simulations of
either systems coupled to external baths\cite{michel08,benenti09,prosen09} or
closed systems prepared in a state out of equilibrium.\cite{langer09,gobert05}

For non-integrable systems, finite-temperature transport is expected to be
diffusive, \cite{castella95,rosch00} which is consistent with the numerical
observation of a vanishing Drude weight for both heat and spin
transport.\cite{hm04,hm03,zotos04,rabson04,alvarez02} While this result has been
accepted for massive phases in the high-temperature regime, the situation for
massless phases of non-integrable systems at low temperatures and in the vicinity
of integrable models\cite{hm03,heidarian07,jung06,jung07a} is less clear, in the
sense that numerical studies yield large ballistic contributions to transport
coefficients in massless phases and on finite systems.\cite{hm07a,heidarian07,footnotelanger10}

Theoretical research into the transport properties of low-dimensional magnets
\cite{hm07a,zotos-review} has been strongly motivated by exciting experimental
results on materials with low-dimensional electronic
structures.\cite{hess07a,sologubenko07a} The large thermal conductivities found
in the spin ladder materials (Sr,Ca,La)$_{14}$Cu$_{24}$O$_{41}$ establish a link
between the thermal conductivity and magnetic
excitations.\cite{hess01,hess04,hess06,sologubenko00} There is also a variety of
spin chain materials with magnetic contributions to the heat conductivity, most
notably Sr$_2$CuO$_3$, SrCuO$_2$
(Ref.~\onlinecite{sologubenko00a,sologubenko01,hlubek10}), and CaCu$_2$O$_3$
(Ref.~\onlinecite{hess07}). Recent successes in sample preparation have resulted in
very clean samples of SrCuO$_2$, exhibiting the largest thermal conductivity so
far observed in low-dimensional quantum magnets.\cite{hlubek10} This has been
interpreted as experimental evidence for ballistic heat transport in clean
Heisenberg chains, where phonons are the main source of external
scattering.\cite{hlubek10}

The magnetic field dependence of the thermal conductivity has been the case of
interest in several experimental studies,
\cite{hess01,sologubenko07,sologubenko08} yet in most of the known materials,
exchange couplings are orders of magnitude larger than the magnetic fields
available in a laboratory. Only recently, the quasi-one-dimensional organic
compound (C$_{5}$H$_{12}$N)$_{2}$CuBr$_{4}$ (Refs.~\onlinecite{patyal90,watson01}) has
gained attention in this context. It has exchange couplings small enough to
allow experimental access to field-driven quantum phase transitions at low
temperatures up to the saturation field. In these experiments, the phase diagram
with respect to temperature and magnetic field has been explored by a large
variety of experimental probes, establishing the presence of a field-induced
gapless phase at low temperatures.\cite{thielemann09,ruegg08,klanjsek08,lorenz08,anfuso08} Recent
measurements of the thermal conductivity of these compounds in external magnetic
fields have been interpreted in terms of the absence of spin mediated heat
transport.\cite{sologubenko09}

The field-dependent thermal transport in the $XXZ$ chain has previously been
addressed with several theoretical approaches,\cite{louis03,sakai05,hm05}
emphasizing the role of magnetothermal corrections to the thermal conductivity
due to the coupling of the spin and the heat current, similar to the Seebeck
effect.\cite{Mahan} The possibility of controlling the heat transport in spin
chains by varying a magnetic field has been addressed in Ref.~\onlinecite{yan09}.
The zero-field transport properties of the dimerized chain have been studied in
Refs.~\onlinecite{hm03,hm07a} while the field-dependence of the thermal Drude weight of
noninteracting, dimerized XX chains has been discussed in Ref.~\onlinecite{orignac03}.

Our present goal is to understand the dependence of the spin and heat
conductivities of dimerized spin chains on external magnetic fields. Within
linear response theory, the frequency-dependent conductivity has two
contributions, a delta peak at zero frequency whose weight is the Drude weight
and a regular part: 
\begin{equation}
\sigma[\kappa](\omega)=D_{\mathrm{s[th]}}\delta(\omega)+\sigma[\kappa]_{\mathrm{reg}}(\omega)
\,. \label{eq:start}
\end{equation}
Using exact diagonalization to evaluate Kubo formulae\cite{Mahan,luttinger64} we
compute these quantities for a value of the spin gap that is comparable to the
one found in (C$_{5}$H$_{12}$N)$_{2}$CuBr$_{4}$
(Refs.~\onlinecite{thielemann09,ruegg08}).

As a main result, we find an increased weight in the
low-frequency regime of both $\sigma(\omega)$ and $\kappa(\omega)$ in
the field-induced phase. This allows for a direct interpretation in terms of
transport channels opened in the vicinity of the Fermi points of the
corresponding Luttinger liquid. These channels lead to a strong enhancement of
the transport coefficients for spin and heat conduction. On the finite systems
that we have access to with exact diagonalization, the main contribution to the
increase of the conductivities at low frequencies and low but {\em
finite} temperatures stems from the Drude weight. In the case of the spin
conductivity this increase is roughly an order of magnitude larger than the one
observed in the regular part, whereas for the thermal conductivity, the picture
is more involved. At zero magnetic field and low temperatures, the bare thermal
Drude weight (without any magnetothermal corrections) dominates the regular
part, yet when increasing the field, the regular part increases more
strongly than the thermal Drude weight. 
Finally, taking into account the magnetothermal corrections, the
increase of the thermal weight with increasing field becomes a {\it decrease},
as expected from the results for the $XXZ$ chain.\cite{hm05}

We further study the dependence of the transport coefficients on the strength of
the dimerization, varying it between the limits of uncoupled dimers and the
Heisenberg chain. We expect that our results are generic for dimerized quasi
one-dimensional systems, while the obvious advantage of working with the
dimerized chain is that with exact diagonalization, we can reach longer chains
than in the case of a ladder. We shall stress that our work is concerned with
the intrinsic transport properties of dimerized systems, whereas for a complete
description of the experimental results, phonons may play an important role, as
has been emphasized in Refs.~\onlinecite{rozhkov05a,chernyshev05,boulat07}.

The paper is organized as follows: First, we introduce the model and briefly
review the ground-state properties. Second, we proceed by summarizing the
necessary framework to compute transport coefficients and conductivities within
linear response theory. Section IV presents the results. We study the Drude
weights in Sec.~\ref{sec:drude}, the frequency-dependent conductivities in
Sec.~\ref{sec:omega} and the spectral weights in Sec.~\ref{sec:ISW}, all as a
function of the magnetic field and temperature. The influence of the strength of
the dimerization is discussed in Sec.~\ref{sec:lambda} and
Sec.~\ref{sec:thermomag} covers magnetothermal effects. Finally, we summarize
our findings in Sec.~\ref{sec:conc}. The dependence of the Drude weights at finite fields on the system size is presented in the Appendix.

\begin{figure}[t]
\includegraphics[width=0.45\textwidth,angle=0]{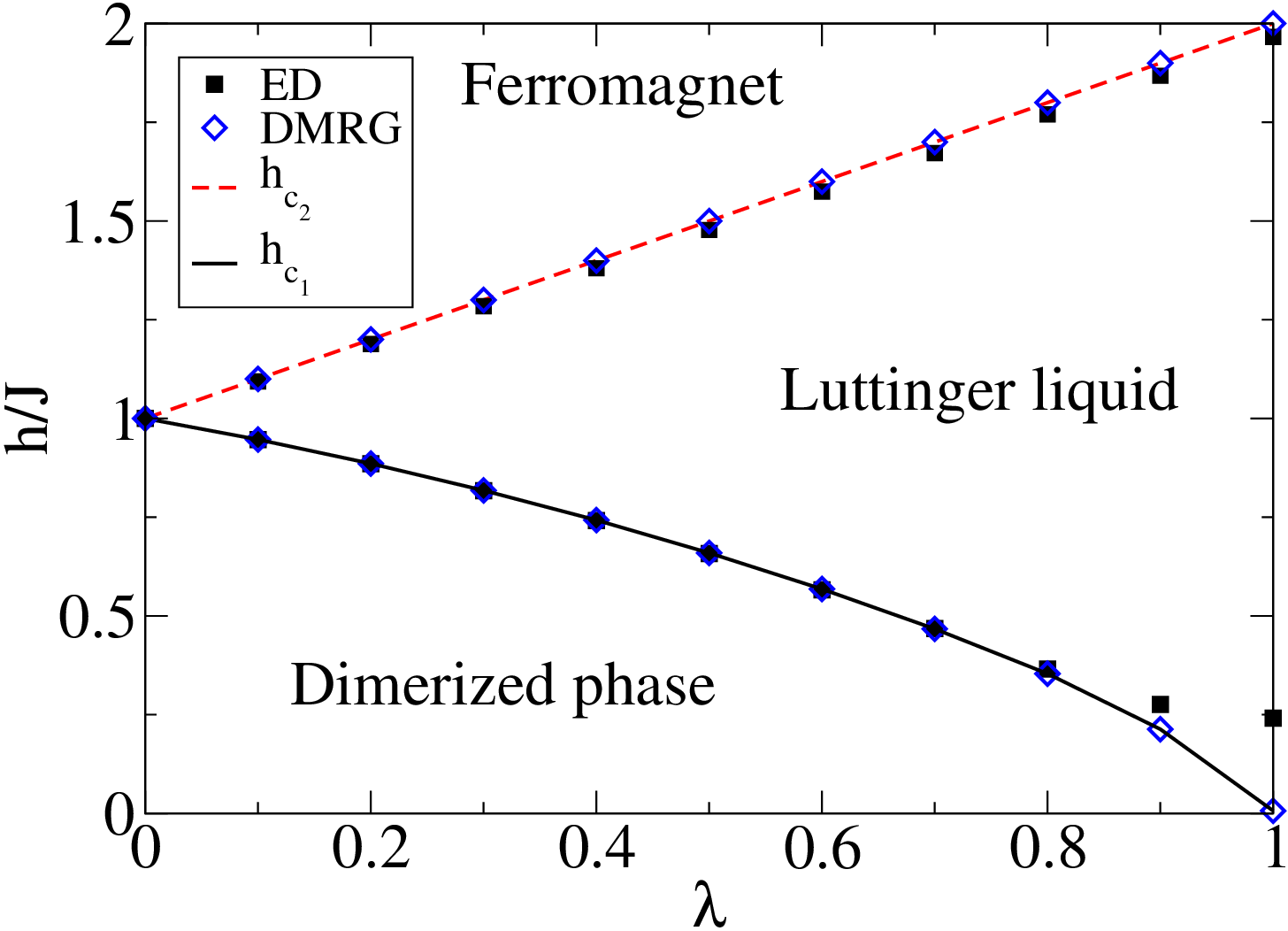}
\caption{(color online) $(h,\lambda)$-plane of the phase diagram of the
dimerized spin chain at temperature $T=0$ (see Refs.~\onlinecite{cabra99,honecker99}). 
The upper (red dashed) line is the saturation
field. The lower (solid black) line corresponds to the spin
gap.\cite{cross79,uhrig99,chitra95,yu99} The critical fields $h_{c_1}$ and
$h_{c_2}$ are obtained from exact diagonalization of chains of length $L=18$ (squares). 
The blue diamonds are DMRG data, extrapolated in the inverse system size.}
\label{fig:phase}
\end{figure}

\section{Model}
We study the spin and thermal conductivity of dimerized spin chains in a homogeneous magnetic field.
The model Hamiltonian for a chain of length $L$ is given by

\begin{equation}
H= \sum_{i=1}^{L}h_i=J \sum_{i=1}^{L} \lbrack \lambda_i \vec{S}_{i}\cdot \vec{S}_{i+1}-hS^z_i\rbrack\,,
\label{eq:model}
\end{equation}

where the dimerization is introduced via: $\lambda_i=\lambda$ for $i$ even and $\lambda_i=1$ for $i$ odd, $\vec{S}=(S^x,S^y,S^z)$, $S_i^{\mu}$ and $\mu=x,y,z$ are the components of a spin-1/2 operator acting on site $i$ and $h$ denotes the magnetic field. We use periodic boundary conditions. The exchange coupling $J$ sets the global energy scale and the model includes the Heisenberg spin chain ($\lambda=1$) and uncoupled dimers ($\lambda=0$) as limiting cases. 

The magnetic phase diagram,\cite{cabra99,honecker99} shown in
Fig.~\ref{fig:phase}, is very similar to the one of the two-leg ladder
system.\cite{chitra96} The upper (red dashed) line is the saturation field. The
lower (solid blue) line corresponds to the spin gap which has been intensely
studied by analytical and numerical means.\cite{cross79,uhrig99,chitra95,yu99}
Since the Heisenberg chain is gapless a careful finite-size scaling for the
lower critical field is necessary. The blue diamonds are
DMRG data for open boundary conditions, extrapolated in inverse system size, which we compare to exact
diagonalization results obtained with $L=18$ sites and periodic boundary
conditions (squares). This illustrates that for $\lambda\lesssim 0.8$, the phase
boundaries exhibit very small finite-size effects; thus, for the system sizes
that we shall use in our exact diagonalization analysis, we are already very
close to the bulk value for the gap. The Luttinger liquid phase here is similar to the Luttinger liquid phase in the XXZ model where the transition from the gapped to the gapless phase is of the commensurate-incommensurate type (see, e.g., Refs.~\onlinecite{cabra98} and \onlinecite{kluemper00}).

\section{Transport coefficients and conductivities}
\label{sec:linres}

\begin{figure*}[t]
\includegraphics[width=0.45\textwidth,angle=0]{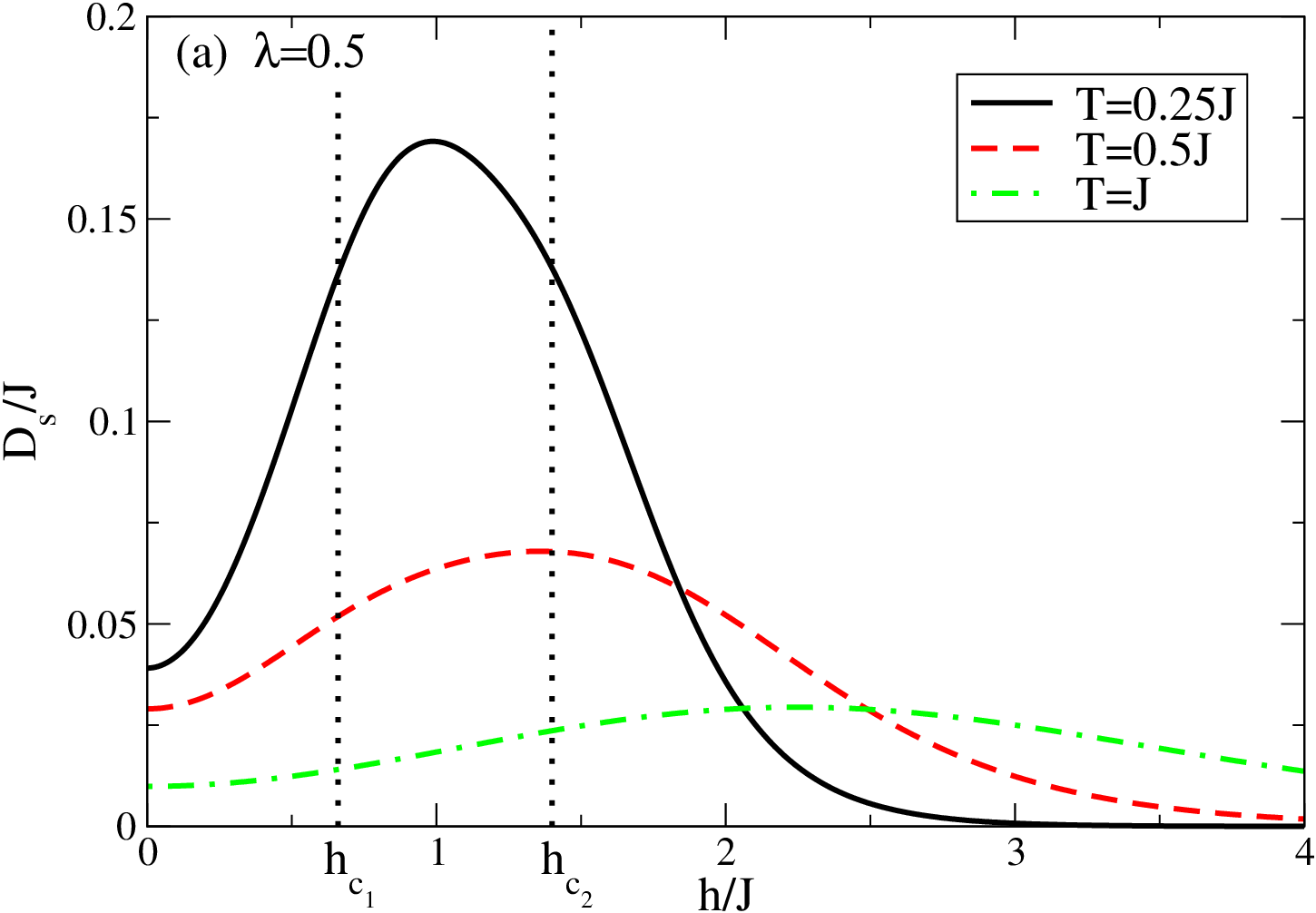}
\includegraphics[width=0.45\textwidth,angle=0]{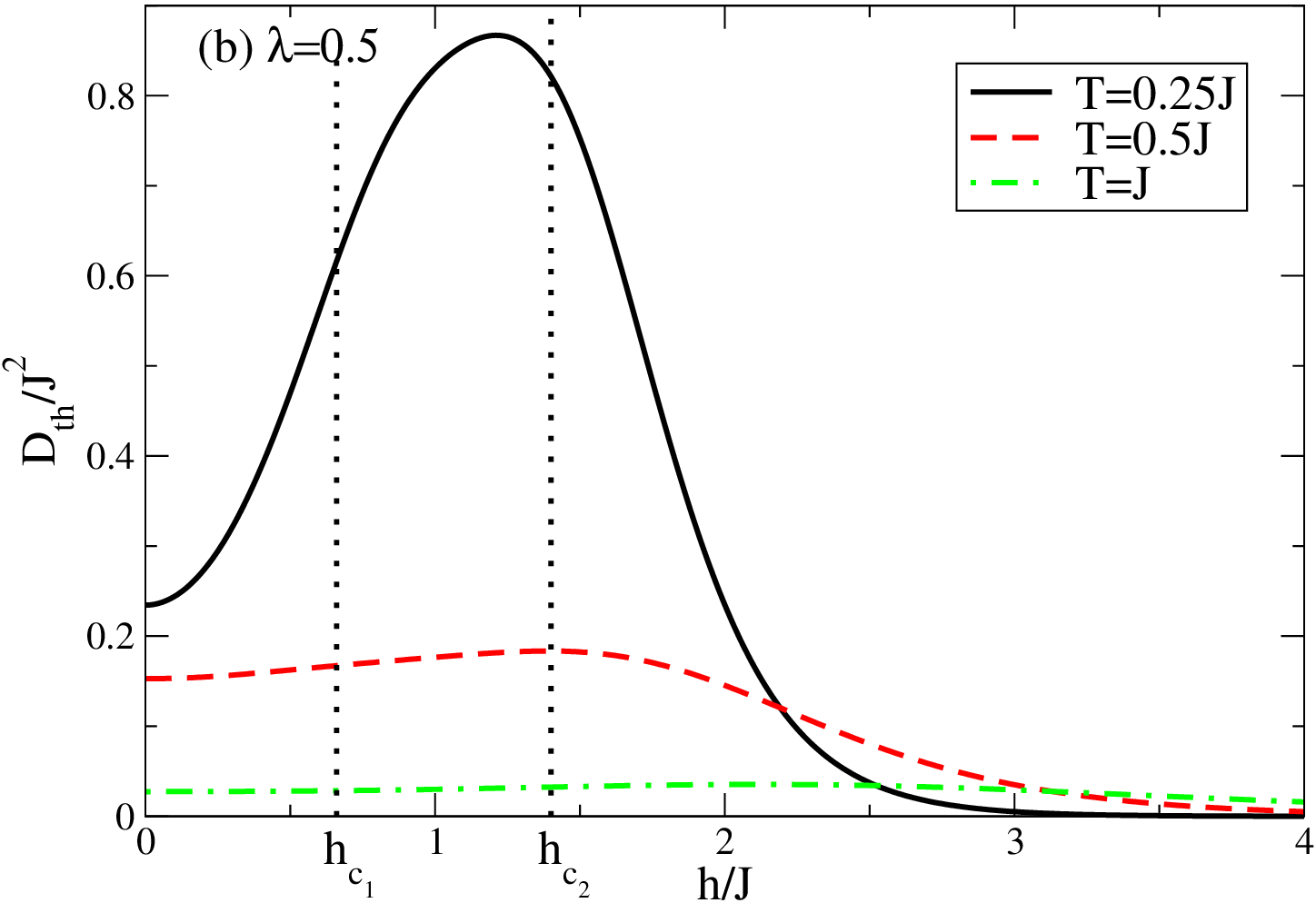}
\caption{(color online) Drude weight for spin (a) and thermal (b) transport at $\lambda=0.5$ and different temperatures for chains of length $L=18$. At low temperatures ($T=0.25J$, solid black lines) we find a significant increase of both quantities in the field-induced phase. At larger temperatures, the peak becomes less pronounced and is located at higher magnetic fields. The black dotted vertical lines are guides to the eye which mark the critical fields $h_{c_1}=0.66J$ and $h_{c_2}=1.4J$.}
\label{fig:drude}
\end{figure*}

Here we summarize the central equations for magnetic transport in the linear
response regime. The expectation values of the spin and thermal currents,
$j_{1}$ and $j_{2}$, are given by\cite{Mahan}
\begin{equation}
\left\langle j_{l}\right\rangle =\sum_{m}L_{lm}f_{m}\,,\label{eq:w1}
\end{equation}
where $f_{1}=\nabla h$ and $f_{2}=-\nabla T$ refer to the magnetic field and
temperature gradients. $L_{lm}$ is the conductivity matrix. $j_{1}$ and
$j_{2}$ can be expressed via the spin and energy currents $j_{\mathrm{s}}$ and $j_{\mathrm{th}}$ by
\begin{equation}
j_{1}=j_{\mathrm{s}},\quad j_{2}=j_{\mathrm{th}}-hj_{\mathrm{s}}\,,\label{eq:w2}
\end{equation}
where
\begin{equation}
j_{\mathrm{s[th]}}=i\sum_{l=1}^{L-1}[h_{l-1},d_{l}]\,.\label{eq:current}
\end{equation}
$h_{l}$ denotes the local energy densities defined by Eq.~\eqref{eq:model} at
zero magnetic field, i.e., $d_{l}=h_l=\lambda_{i}S_{i}^{\mu}S_{i+1}^{\mu}$ for
heat transport, while
$d_{l}=S_{l}^{z}$ for spin transport. The conductivities satisfy
\begin{equation}
\sigma_{lm}(\omega)\equiv\mathrm{Re\,}L_{lm}\left(\omega\right)=D_{lm}\delta(\omega)+\sigma_{\mathrm{reg},lm}(\omega)\label{eq:w3}
\end{equation}
with
\begin{equation}
D_{lm}=\frac{\pi\beta^{r+1}}{ZL}\sum_{\stackrel{n,o}{E_{n}=E_{o}}}e^{-\beta E_{n}}\langle n|j_{l}|o\rangle\langle o|j_{m}|n\rangle\,,\label{eq:w4}
\end{equation}
where $r=0(1)$ for $m=1(2)$ and
\begin{eqnarray}
\lefteqn{\sigma_{\mathrm{reg},lm}=\frac{\pi\beta^{r}}{ZL}\frac{1-e^{-\beta\omega}}{\omega}\sum_{\stackrel{n,o}{E_{n}\neq E_{o}}}e^{-\beta E_{n}}}\nonumber \\
 & & \hphantom{aaaaa}\times\langle n|j_{\mathrm{l}}|o\rangle\langle o|j_{\mathrm{m}}|n\rangle\delta(\omega-\Delta E)\,.\label{eq:w5}
\end{eqnarray}
$\beta=1/T$ is the inverse temperature, $|n\rangle$ and $E_{n}$ are the
eigenstates and energies of $H$, $Z=\sum_{n}e^{-\beta E_{n}}$ denotes the
partition function, and $\Delta E=E_{o}-E_{n}$. Next we define
\begin{equation}
\left.\begin{array}{l}
\sigma\left(\omega\right)\\
\kappa\left(\omega\right)\\
\sigma_{\mathrm{th,s}}\left(\omega\right)\end{array}\right\} \equiv\left\{ \begin{array}{l}
\sigma_{11}\left(\omega\right)\\
\left.\sigma_{22}\left(\omega\right)\right|_{j_2\rightarrow j_{\mathrm{th}}}\\
\left.\sigma_{21}\left(\omega\right)\right|_{j_2\rightarrow j_{\mathrm{th}}}\end{array}\right.\,,\label{eq:w7}
\end{equation}
and the Drude weights $D_{\mathrm{s}}$, $D_{\mathrm{th}}$, and $D_{\mathrm{th,s}}$ of $\sigma(\omega)$,
$\kappa(\omega)$, and $\sigma_{\mathrm{th,s}}(\omega)$, respectively. 
The latter set of quantities corresponds to a choice for the currents alternative to Eq.~(\ref{eq:w2}), namely
$j_1=j_{\mathrm{s}}$ and $j_2=j_{\mathrm{th}}$. Note that
$TD_{12(\mathrm{s,th})}=D_{21(\mathrm{th,s})}$. As in
Refs.~\onlinecite{hm04,hm03,hm05,hm07a} we will evaluate
$\sigma\left(\omega\right)$, $\kappa\left(\omega\right)$, and
$\sigma_{\mathrm{th,s}}\left(\omega\right)$ using exact diagonalization. From these
quantities all transport coefficients $\sigma_{lm}(\omega)$ at finite fields can
be obtained using Eqs.~(\ref{eq:w2}), (\ref{eq:w4}), and (\ref{eq:w5}). The
Drude weight $K_{\mathrm{th}}$ for \emph{purely} thermal transport accounts for
the situation of \emph{no} spin current, i.e., at $\langle j_{\mathrm{s}}\rangle=0$. It
is obtained analogously to the Seebeck effect as
\begin{equation}
K_{\mathrm{th}}=D_{\mathrm{th}}-\frac{D_{\mathrm{th,s}}^{2}}{TD_{\mathrm{s}}}\,,\label{eq:thmag1}
\end{equation}
or equivalently, using Eqs.~(\ref{eq:w2}) and (\ref{eq:w4}): 
$K_{\mathrm{th}}=D_{22}-D_{21}^{2}/(TD_{11})$. 
We will study the field dependence of the magnetothermal coupling in
Sec.~\ref{sec:thermomag}.

For the spin Drude weight, an alternative expression fully equivalent to
Eq.~(\ref{eq:w4}) exists\cite{kohn64}
\begin{equation}
D_{\mathrm{s}}=\frac{\pi}{ZL}[\langle-\hat{T}\rangle-2\sum_{\stackrel{n,o}{E_{n}\neq E_{o}}}e^{-\beta E_{n}}\frac{|\langle o|j_{\mathrm{s}}|n\rangle|^{2}}{\Delta E}]\,,\label{eq:w6}
\end{equation}
where $\langle\hat{T}\rangle$ denotes the kinetic energy. 
This expression will be used to
evaluate $D_{\mathrm{s}}$ in this paper since it gives the correct contribution to the optical sum rule Eq.~\eqref{eq:opt} on finite systems at low temperatures (see the discussion below).\cite{hm05}

Finally, we define the integrated spectral weights
\begin{equation}
I_{\mathrm{s[th]}}(\omega):=\int_{-\omega}^{\omega}\sigma[\kappa](\omega)d\omega=D_{\mathrm{s[th]}}+2\int_{0^{+}}^{\omega}\sigma[\kappa]_{\mathrm{reg}}(\omega)\,,
\end{equation}
and $I_{\mathrm{s[th]}}^{0}\equiv I_{\mathrm{s[th]}}(\infty)$. For the spin
conductivity one obtains the optical sum rule\cite{shastry90}
\begin{equation}
I_{\mathrm{s}}^{0}:=\int_{-\infty}^{\infty}\sigma(\omega)d\omega=\frac{\pi}{L}\langle-\hat{T}\rangle\,.\label{eq:opt}
\end{equation}
The r.h.s. of the corresponding sum rule for thermal transport \cite{shastry06} at finite temperatures
depends on the model and the choice for the local energy density.

\begin{figure*}[t]
\includegraphics[width=0.45\textwidth,angle=0]{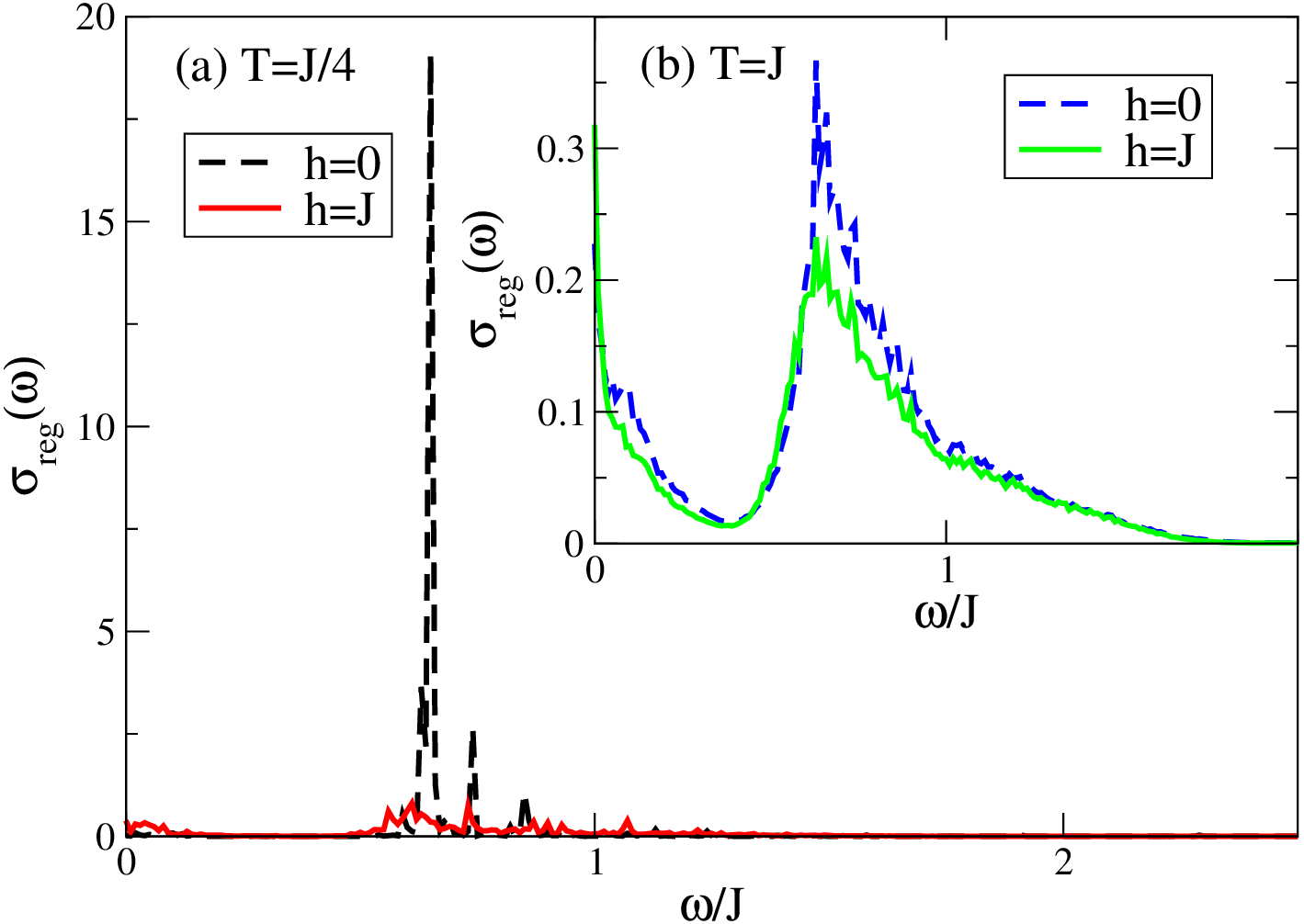}
\includegraphics[width=0.45\textwidth,angle=0]{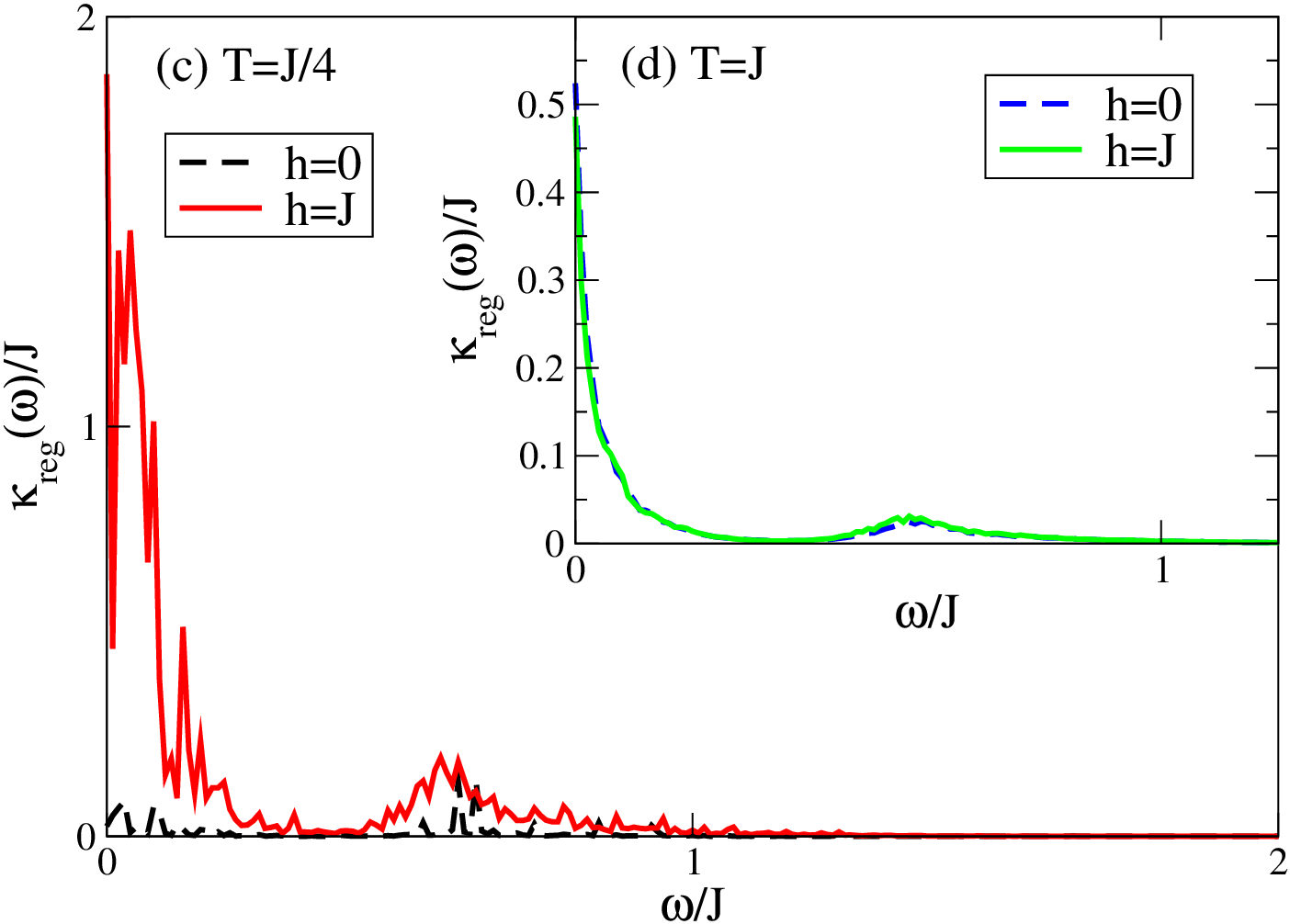}
\caption{(Color online) Regular part of the dynamical conductivities $\sigma(\omega)$, $\kappa(\omega)$ as a function of frequency for different magnetic fields and temperatures. (a) shows the spin conductivity at low temperatures $(T=0.25J)$ and zero magnetic field (dashed black line) as well as $h=J$ (solid red line) for a chain with $L=18$ spins and a dimerization of $\lambda=0.5$. In the presence of a field $h=J$, $\sigma_{\mathrm{reg}}(\omega)$ increases in the gap, but strongly decreases at higher frequencies. The inset (b) shows the same at a higher temperature $(T=J)$ where the effect of $h$ is less strong. (c) and (d) show $\kappa_{\mathrm{reg}}(\omega)$ of the same system. (c) For low temperatures, the magnetic field gives rise to an increase of $\kappa_{\mathrm{reg}}(\omega)$ over the whole spectrum (solid red line). The inset (d) shows $\kappa_{\mathrm{reg}}(\omega)$ at higher temperatures $(T=J)$, where $\kappa_{\mathrm{reg}}(\omega)$ is a smooth curve and barely influenced by the field.}
\label{fig:conduct}
\end{figure*}

\begin{figure*}[tb]
\includegraphics[width=0.45\textwidth,angle=0]{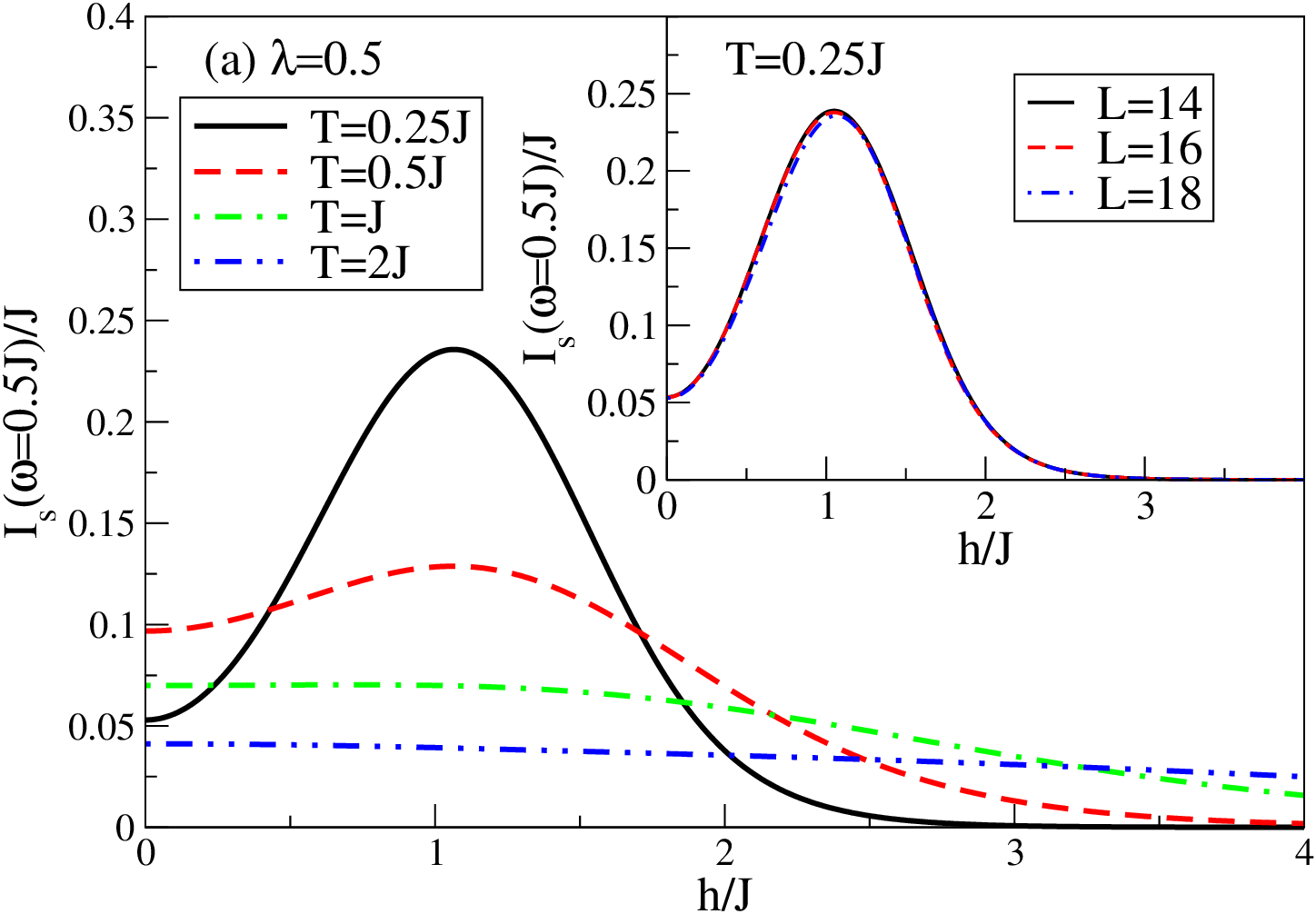}
\includegraphics[width=0.45\textwidth,angle=0]{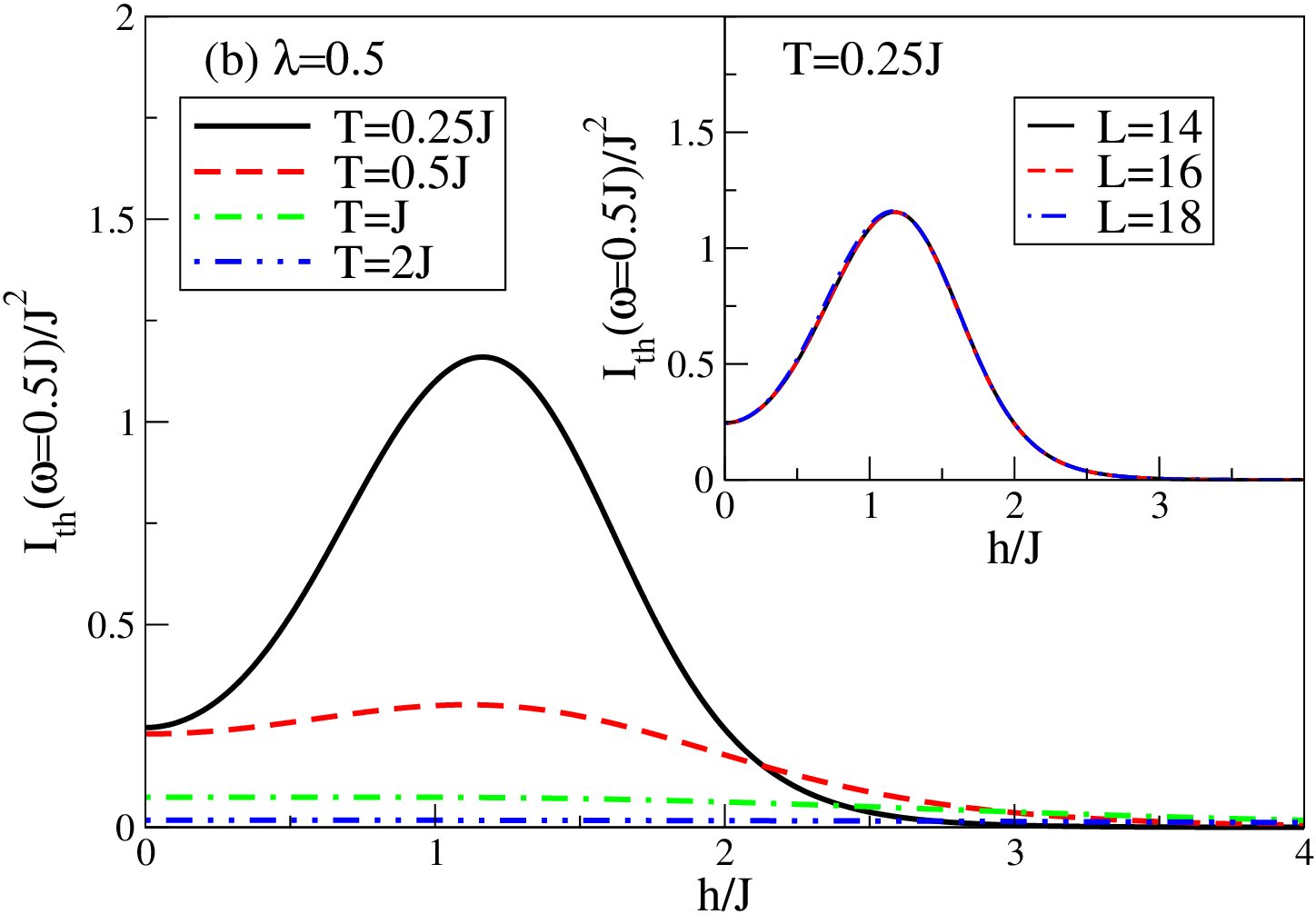}
\includegraphics[width=0.45\textwidth,angle=0]{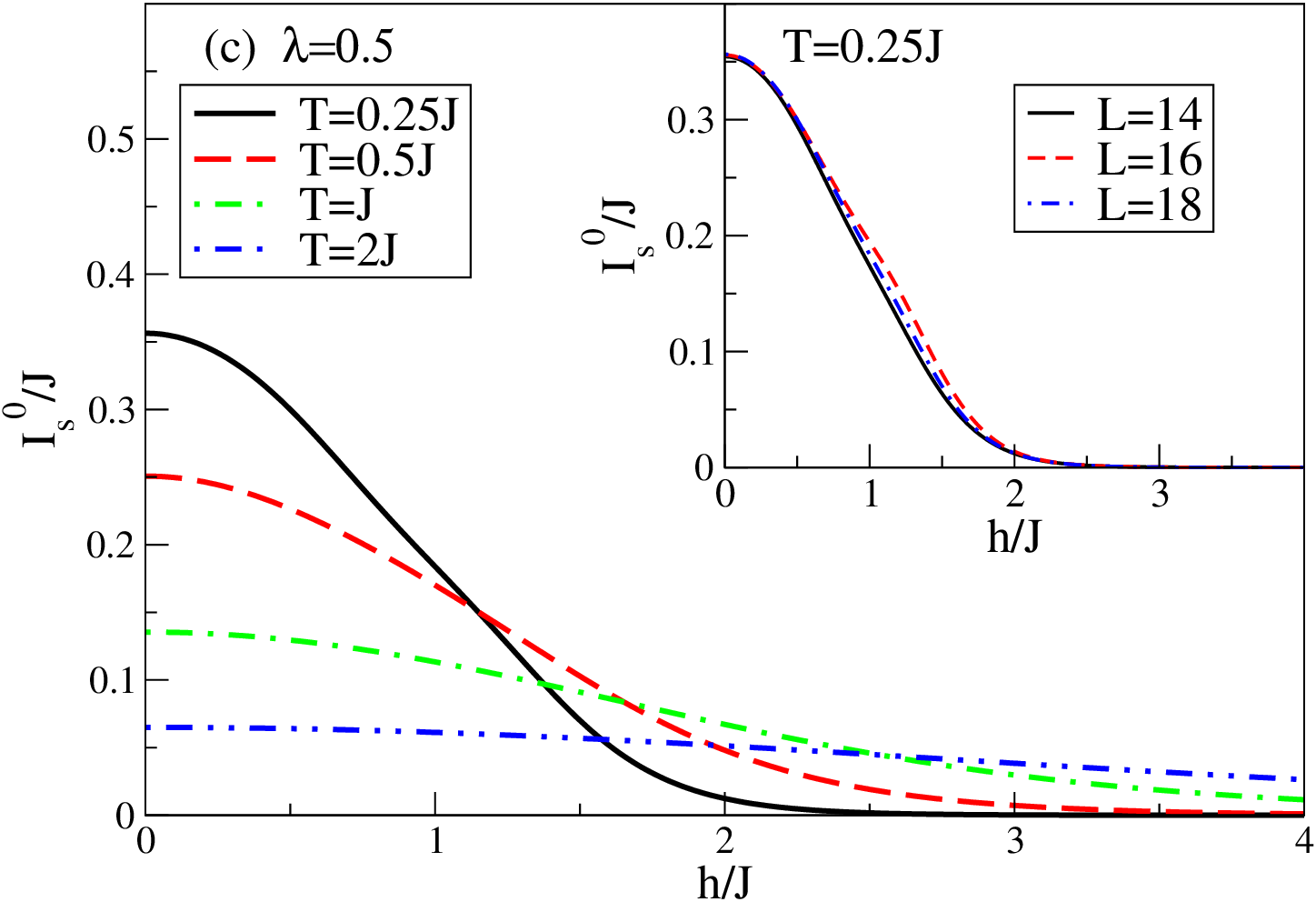}
\includegraphics[width=0.45\textwidth,angle=0]{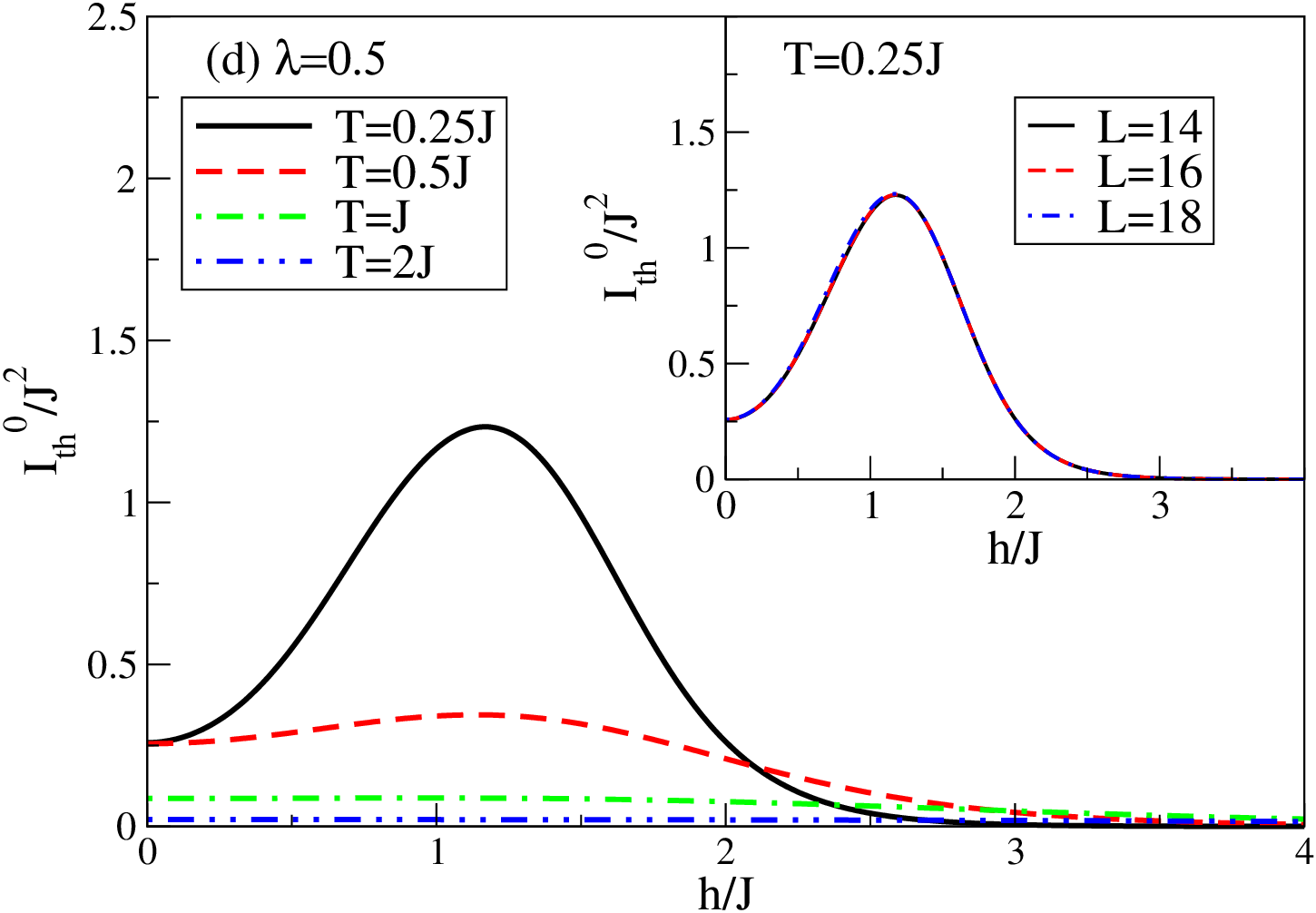}
\caption{(Color online) (a) Spectral weight $I_{\mathrm{s}}(\omega)$ for spin transport integrated up to $\omega=0.5J$ as a function of magnetic field $h$ for $L=18$ spins and at different temperatures. (b) Spectral weight $I_{\mathrm{th}}(\omega)$ for heat transport integrated up to $\omega=0.5J$. At low temperatures, there is a large peak in both quantities indicating the enhancement of transport in the field-induced gapless phase. As temperature increases this effect vanishes up to the point where the magnetic field causes the 
decrease of spectral weight in the gap. For heat transport, the magnetic field has almost no effect at large temperatures ($T=J,2J$). (c), (d) When extending the upper bound on the frequency $\omega$ to infinity, the effect of the magnetic field on spin transport is hidden under the large contributions of the regular part above the gap (c), while $I^0_{\mathrm{th}}$ behaves similar to $I_{\mathrm{th}}(\omega=J/2)$ (d). The inset in each panel illustrates the small dependence on the system size at low temperatures.}
\label{fig:I05}
\end{figure*}

\section{Results}

We present our results for the transport coefficients of dimerized spin chains, starting with the dimerization strength $\lambda=0.5$, yielding a gap close to the one of the experimental system. We separately analyze the Drude weights (Sec.~\ref{sec:drude}) and the conductivities (Sec.~\ref{sec:omega}) and then combine them to get the spectral weight (Sec.~\ref{sec:ISW}), all as a function of temperature and magnetic field. As a result, we find that the dominant effect of applying a magnetic field for this fixed value of $\lambda$ is an enhancement of the low-frequency spin as well as heat conductivity. In Sec.~\ref{sec:lambda}, we analyze this enhancement at different strengths of the dimerization. Section~\ref{sec:thermomag} discusses the magnetothermal coupling. 
\subsection{Drude weights as a function of $T$ and $h$}
\label{sec:drude}
We start focusing on $\lambda=0.5$, resulting in a value of the spin gap
$\Delta=0.66$ in units of $J$. One of the two contributions to the integrated
spectral weights is the Drude weight in
Eq.~\eqref{eq:start}. Figure~\ref{fig:drude} shows the spin Drude weight
$D_{\mathrm{s}}$ and the thermal Drude weight $D_{\mathrm{th}}$ as a function of the
magnetic field for $T/J=0.25,0.5,1$ and chains consisting of $L=18$ spins. At
low temperatures ($T=0.25J$, solid black lines), we find a strong enhancement of
both quantities between the critical fields $h_{c_1}=0.66J$ and
$h_{c_2}=1.48J$. Increasing the temperature weakens this feature significantly,
especially for thermal transport. This feature is addressed quantitatively in the discussion of the spectral weights (Sec.~\ref{sec:ISW}). It is also worth noting that the maximum in
the Drude weights $D_{\mathrm{s[th]}}=D_{\mathrm{s[th]}}(h)$ moves to higher fields as the temperature increases. At low
temperatures, the question arises whether there are any features in
$D_{\mathrm{s}}=D_{\mathrm{s}}(h)$ at the critical field, i.e., whether
the finite systems we investigate exhibit remnants of the quantum critical
behavior to be expected at $h_{c_1}$. We shall comment on this point in
Sec.~\ref{sec:lambda}, where we will discuss the $\lambda$-dependence of the
Drude weights.

\begin{figure}[t]
\includegraphics[width=0.45\textwidth,angle=0]{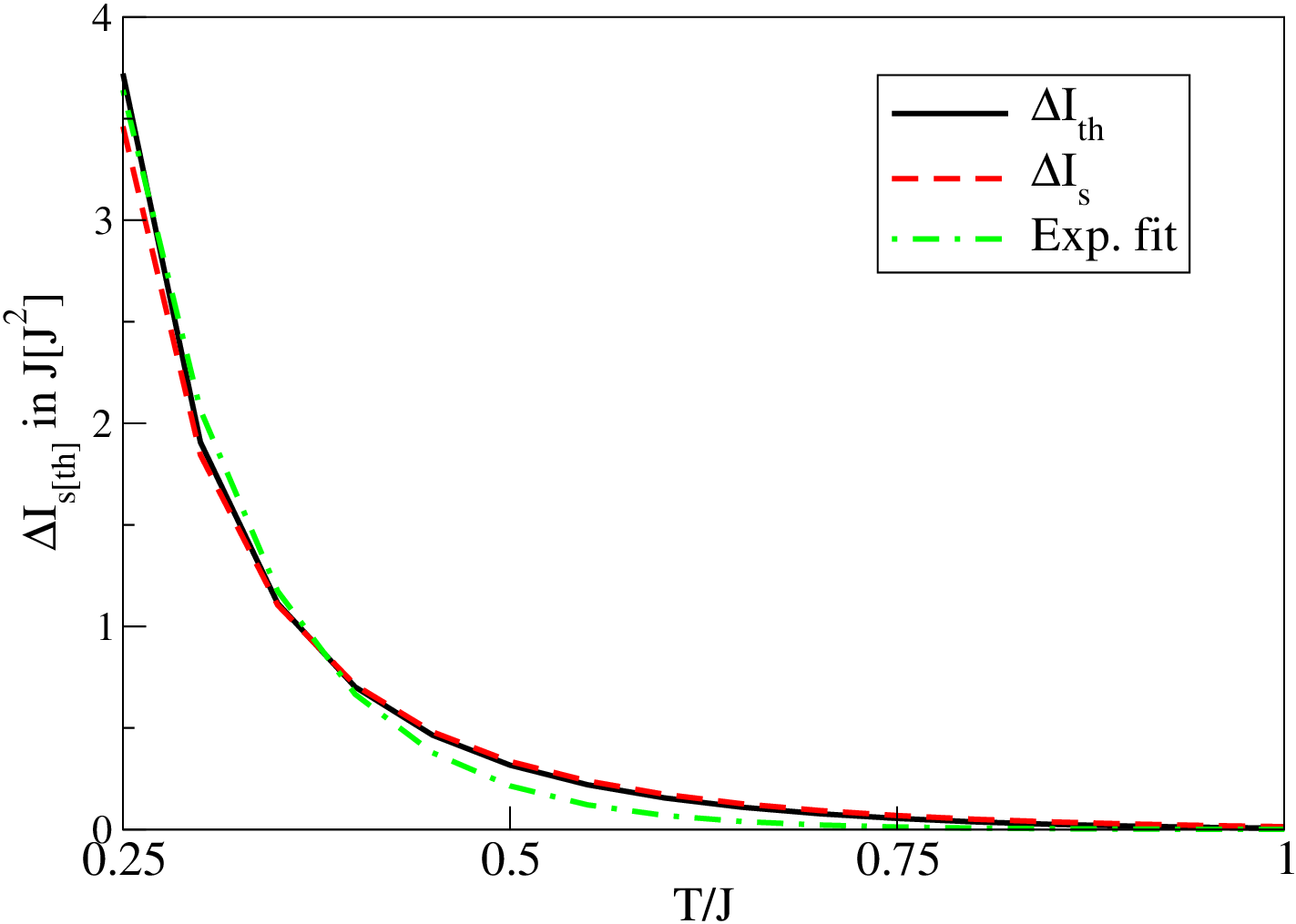}
\caption{(color online) Relative enhancement $\Delta I_{\mathrm{s[th]}}$ of the spectral weights as a function of temperature $(L=14)$. The curves for thermal transport (solid black line) and spin transport (dashed red line) almost coincide and decay slightly slower than exponentially ($\Delta I_{\mathrm{s[th]}}\sim \mbox{exp}(-cT)$, green dash-dotted line) as the temperature increases.}
\label{fig:enh}
\end{figure}

\begin{figure*}[tb]
\includegraphics[width=0.45\textwidth,angle=0]{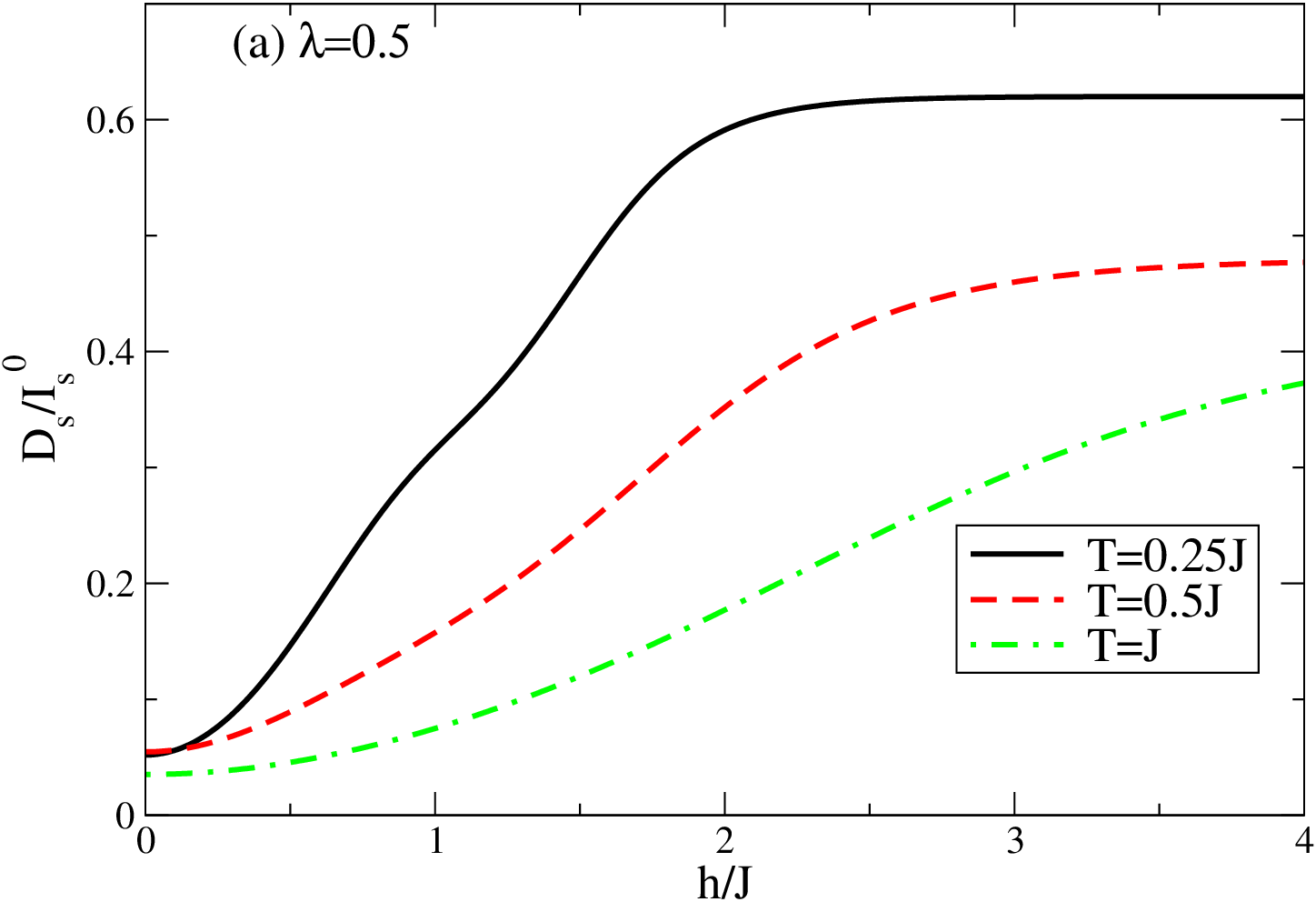}
\includegraphics[width=0.45\textwidth,angle=0]{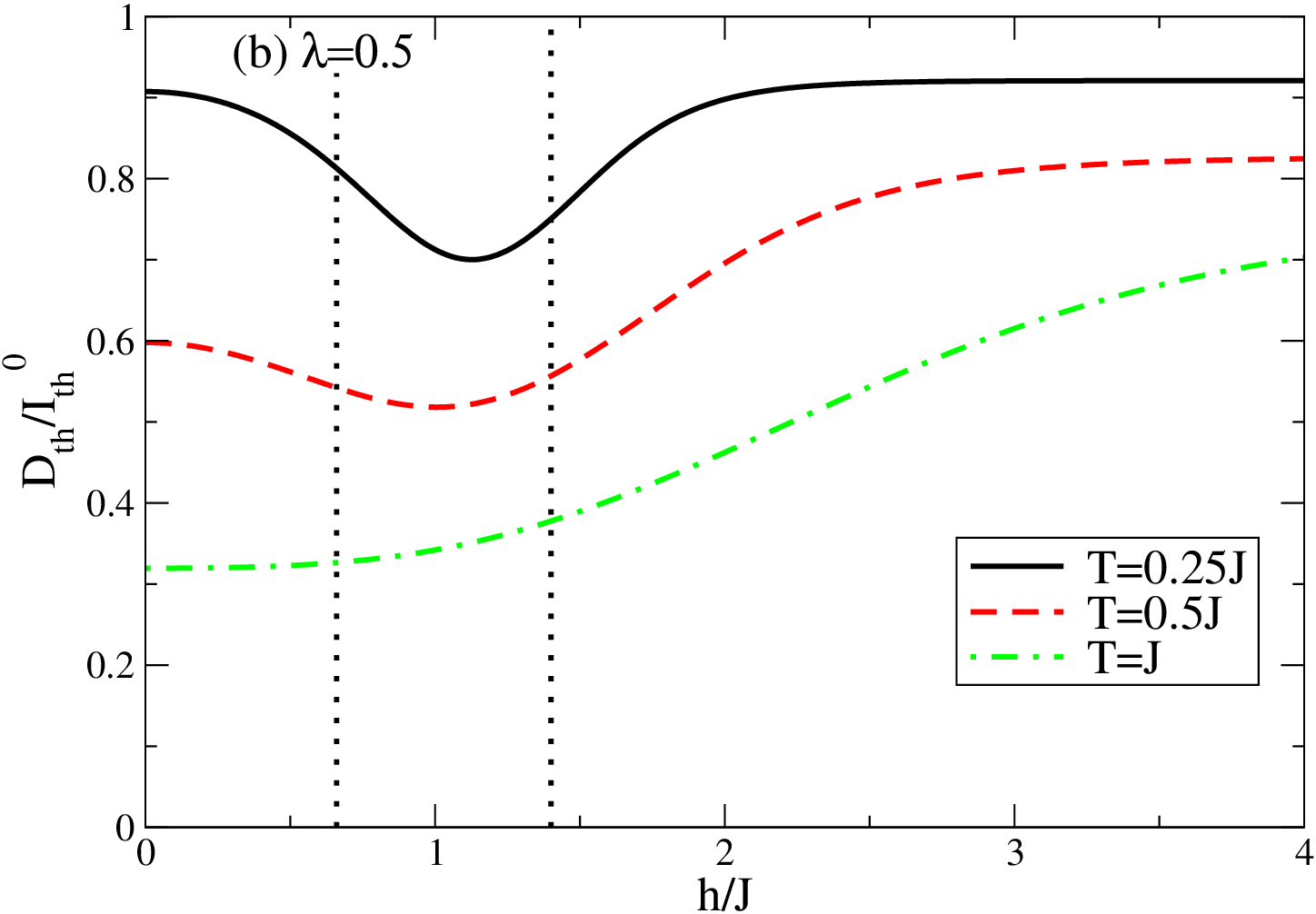}
\caption{(Color online) Drude weights normalized by the total spectral weight $I^0_{\mathrm{s[th]}}$ as a function of magnetic field. (a) In the case of spin transport, the Drude weight is a small contribution at zero field, but increases with the magnetic field up to a saturation value that strongly depends on temperature. (b) For heat transport the Drude weight is the dominant contribution even at zero field and low temperatures. At low temperatures the normalized Drude weight has a minimum as the field increases. At $h\approx 2J$, it reaches its final value. Such a minimum is not visible at $T=J$. The black dotted vertical lines are guides to the eye which mark the critical fields $h_{c_1}=0.66J$ and $h_{c_2}=1.4J$.
}
\label{fig:DrudeI0}
\end{figure*}

\subsection{Dynamical conductivities}
\label{sec:omega}
The second contribution to the spectral weight is the regular part [Eq.~\eqref{eq:w5}] of the dynamical conductivity. By inspecting the spectral representations, one realizes that the field can only enter via the Boltzmann factors, thus a magnetic field can only enhance or reduce existing weight but does not change the position of the poles $\omega=\Delta E$. Note that at zero magnetic field, the weight in the gap is suppressed at low temperatures as the system approaches its dimerized ground-state and can be tuned to finite values by increasing either the magnetic field or temperature.\cite{hm07a} 

Our numerical results are depicted in Fig.~\ref{fig:conduct}. At zero magnetic
field and low temperatures, there is almost no weight in the regular part of
$\sigma(\omega)$ [dashed line in Fig.~\ref{fig:conduct}(a)] below a value of
$\omega$ approximately corresponding to $h_{c_1}$ (compare
Ref.~\onlinecite{hm07a} for the case of $\lambda=0.1$ and $h=0$) beyond which
the dominant peak is located. Turning on a magnetic field larger than
$h_{c_1}$ influences the curve drastically. The major portion of the weight
still lies above the gap but is much smaller and without significant peaks
[solid line in Fig.~\ref{fig:conduct}(a)]. In addition to the overall reduction we find 
spectral weight at very small frequencies which is not the case at zero
field. Going to higher temperatures [Fig.~\ref{fig:conduct}(b)] results in a
smooth curve due to thermal excitations, while the influence of the magnetic
field is much weaker and does only change the numerical values
without modifying the structure.

In the case of thermal transport [Fig.~\ref{fig:conduct}(c)], the basic
structure is different and the role of the magnetic field at low temperatures is
even more important. Without a field (dashed line) a significant amount of spectral weight is
found around $\omega=0$ while another important contribution is at the lower
edge of the one-triplet band, similar to the spin conductivity. Switching on a
magnetic field larger than $h_{c_1}$ (solid line) enhances the heat conductivity
strongly, especially at low frequencies. Increasing the temperature up to $T=J$ smoothens the curve
while completely suppressing the influence of the magnetic field. 

Concentrating
on the regular parts alone, one would conclude that both spin and heat transport
are not influenced by the magnetic field at high temperatures. At low
temperatures the weight in $\sigma_{\mathrm{reg}}(\omega)$ is strongly reduced by the magnetic
field yet comes along with an increase in the weight at low frequencies, which is the effect we are interested in.
By contrast, $\kappa_{\mathrm{reg}}(\omega)$ is
enhanced by the field at low temperatures at all frequencies, with most of this 
increase originating from the low-frequency range $\omega < h_{c_1}$. At higher
temperatures, both effects are suppressed. To understand the interplay of the
Drude weights and the regular parts one has to study the spectral weight.

\begin{figure*}[t]
\includegraphics[width=0.45\textwidth,angle=0]{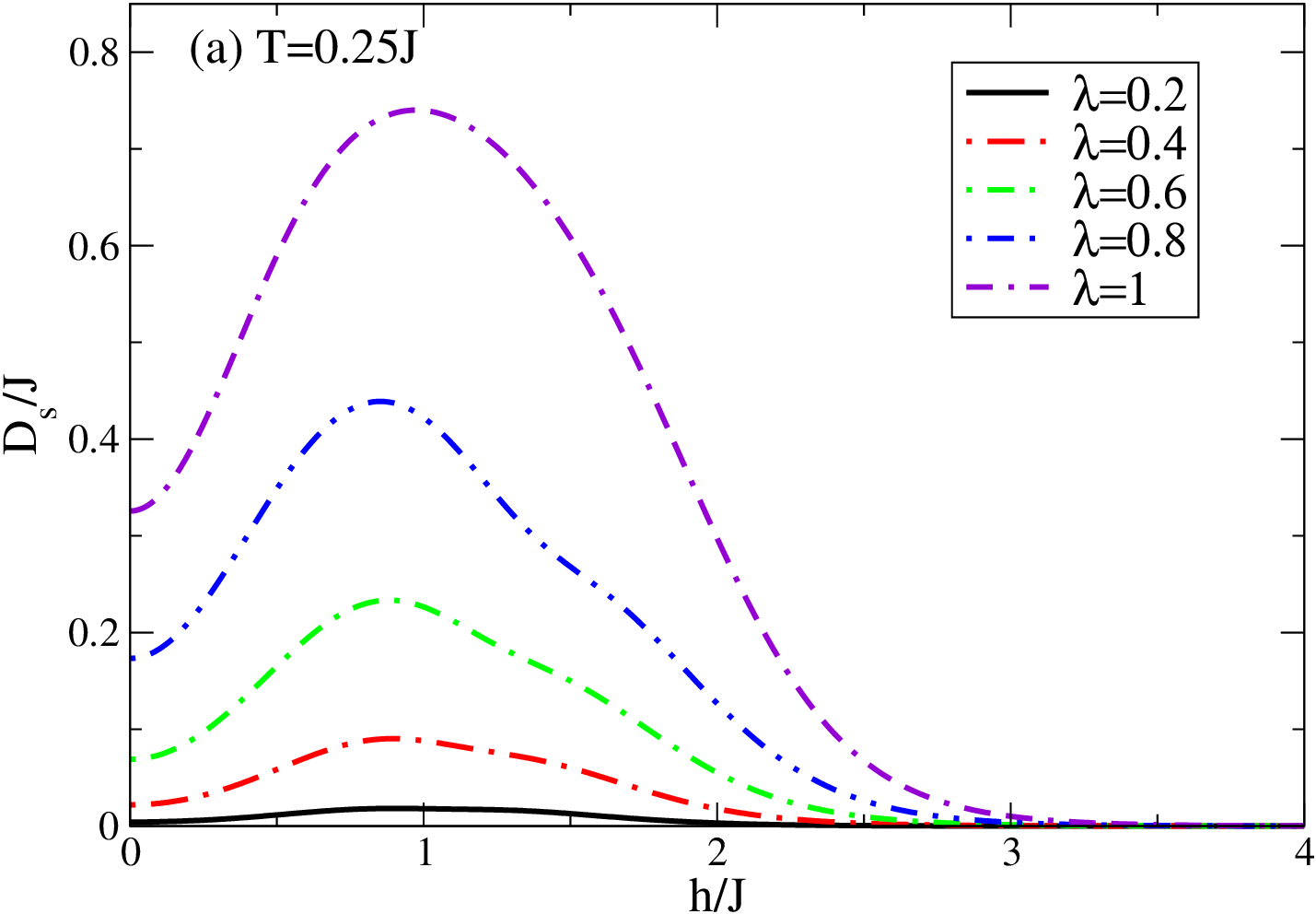}
\includegraphics[width=0.45\textwidth,angle=0]{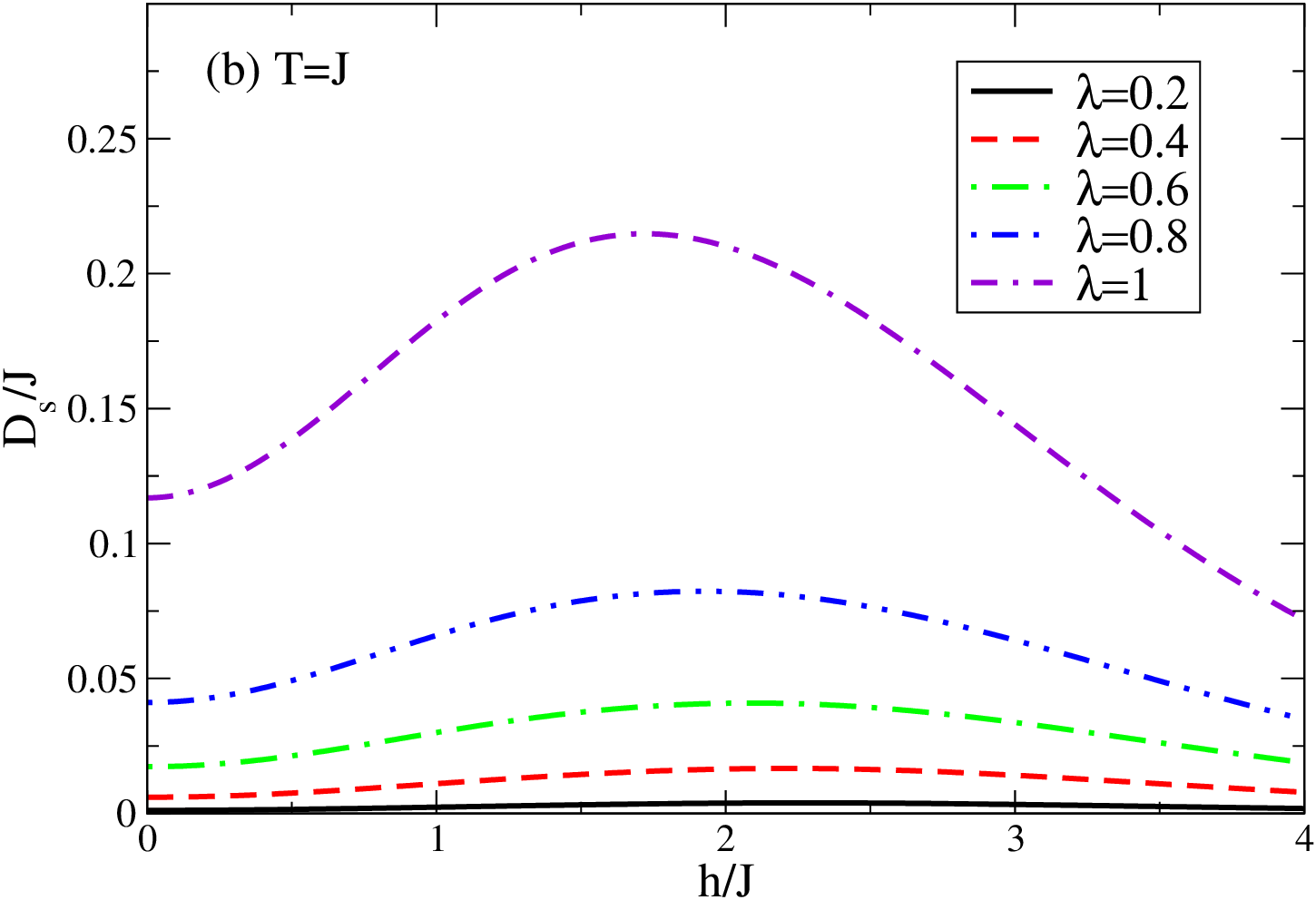}
\caption{(Color online) Spin Drude weight as a function of magnetic field $h$ for a system of length $L=16$ and different values of $\lambda$. (a) At low temperatures ($T=0.25J$) the spin Drude weights show a significant peak between the critical fields (compare Fig. \ref{fig:phase} at $T=0$). (b) For larger temperatures ($T=J$) the peak is less pronounced and moves to larger fields. In both cases increasing $\lambda$ gives rise to a strong increase of the numerical values but the position of the peak moves only slightly towards smaller fields. The qualitative field dependence is almost independent of $\lambda$. For all values of $\lambda$ the peak is located at marginally smaller fields than for the Heisenberg chain [long dashed lines in (a)].
}
\label{fig:drudelam}
\end{figure*}

\subsection{Field-dependence of the spectral weight}
\label{sec:ISW}
Now that we have shown how the transport enhancement takes place via increasing weight at low frequencies we study the integrated spectral weight $I_{\mathrm{s[th]}}(\omega)$. 
Figure~\ref{fig:I05} shows our results for $I_{\mathrm{s}}(\omega=0.5J)$ and $I_{\mathrm{th}}(\omega=0.5J)$ as well as the total spectral weights $I_{\mathrm{s[th]}}^0$. 
We have chosen the upper bound of integration to be $\omega=0.5J$ to measure how the weight in the gap of the system behaves as a function of temperature and field. 
Our findings in the case of low temperature and $\lambda=0.5$ are the following: For both spin and heat transport, we find a strong enhancement of transport in the field-induced gapless phase at low temperatures $T<h_{c_1}$. 
As temperature increases, the field dependence becomes less pronounced. 
From Fig.~\ref{fig:I05}, it is obvious that the total weight in the spin conductivity, i.e., $I^0_{\mathrm{s}}\propto\langle -\hat T\rangle$ typically decreases as a function
of increasing field, due to the increasing weight of strongly polarized states.
Therefore, the enhanced low-frequency weight seen in both the Drude weight and the regular part has to be accompanied by a decrease of spectral weight at 
higher frequencies in order to satisfy the optical sum rule Eq. \eqref{eq:opt}, consistent with our discussion of Fig.~\ref{fig:conduct}(a).

In the case of thermal transport, the influence of temperature on
$I_{\mathrm{th}}^0$ is much more drastic: the maximum in $I_{\mathrm{th}}^0$
seen at $T=0.25J$ is quickly washed out at temperatures $T>h_{c_1}$. By
contrast, for spin transport and in $I_{\mathrm{s}}^0$, the interesting features
are hidden at all temperatures under the large field-induced decrease of the
regular part as $\omega$ approaches the gap [Fig.~\ref{fig:I05}(c)], 
whereas in the case of thermal transport, the low-frequency behavior dominates
the field dependence of $I_{\mathrm{th}}^0$ at $T<h_{c_1}$
[Fig.~\ref{fig:I05}(d)]. For $T>0.25J$, the finite-size effects of the integrated weights become very small, which is shown in the insets of Fig.~\ref{fig:I05}, illustrating that this quantity is very robust. 
In other words, the maximum of $I_{\mathrm{s[th]}}(\omega=0.5J)$ in the window $h_{c_1}< h< h_{c_2} $ is
stable against a variation in $L$.

The field-dependent enhancement of $I_{\mathrm{s[th}]}(\omega=0.5J)$ in Fig.~\ref{fig:I05} (a) and (b) $(T=0.25J)$ is well described by $I_{\mathrm{s[th]}}\sim h^2$ for $h\leq h_{c_1}$. Furthermore we present 
$$\Delta I_{\mathrm{s[th]}}=\frac{\mathrm{max}_h(I_{\mathrm{s[th]}}(\omega=0.5J,h))-I_{\mathrm{s[th]}}(\omega=0.5J,h=0)}{I_{\mathrm{s[th]}}(\omega=0.5J,h=0)}$$ 
in Fig.~\ref{fig:enh} as a function of temperature for $L=14$ spins. The data are already converged with respect to the system size for the temperatures shown. While the field-dependence of $I_{\mathrm{th}}(\omega=0.5J)$ and $I_{\mathrm{s}}(\omega=0.5J)$ is only qualitatively similar, $\Delta I_{\mathrm{th}}$ (black solid line) and $\Delta I_{\mathrm{s}}$ (red dashed line) almost coincide. The green dash-dotted line is an exponential decay fitted to $\Delta I_{\mathrm{th}}$ to illustrate that the decay is slightly slower than exponential for spin as well as for heat transport.

Regarding the temperature dependence of the thermal spectral weight at lower temperatures $(T<0.25J)$ we expect an exponential suppression at low temperatures in gapped phases\cite{orignac03,hm03,hm02} ($h<h_{c_1}$ or $h>h_{c_2}$). For the Luttinger Liquid phase ($h_{c_1}<h<h_{c_2}$), we expect $I_{\mathrm{s[th]}}=I_{\mathrm{s[th]}}(T)\sim T$ based on previous work on the temperature dependence of the thermal drude weights of the XXZ spin-1/2 chain in a magnetic field.\cite{hm05}
At zero magnetic field we indeed find an exponentially suppressed weight converged with respect to the systems size for $T<0.25J$, but in the gapless phase ($h=J$) the finite-size effects do not allow for a definite conclusion on the temperature dependence.

To clarify the role of the Drude weight on finite systems in this context, it is plotted
in Fig.~\ref{fig:DrudeI0} as a fraction of the corresponding total
spectral weight. In the case of spin transport [Fig.~\ref{fig:DrudeI0} (a)], the
Drude weight contributes very little to the total weight but its significance
increases monotonously with the field. $D_{\mathrm{s}}/I^0_{\mathrm{s}}$ saturates
at a finite value above the saturation field. These observations hold for all
temperatures studied ($T/J=0.25,0.5,1$), only the value at saturation and the
field necessary for reaching it depends on temperature. For instance, at
$T=0.25J$ (solid black line) the Drude weight accounts for $60\%$ of the total
weight above $h=2J$. At $T=0.5J$ (dashed red line),
$D_{\mathrm{s}}/I^0_{\mathrm{s}}$ saturates at $h=3.5J$ at a value of
$48\%$. However, Fig.~\ref{fig:DrudeI0} should not be interpreted in terms of a
large absolute value of the Drude weight above saturation
in the thermodynamic limit, neither at zero or finite temperatures. At zero temperature, for $N\to \infty$ and in the two gapped phases
$h<h_{c_1}$ and $h>h_{c_2}$, we expect the spin Drude weight to vanish, according to Kohn's
reasoning.\cite{kohn64} As far as the gapless phase is concerned, the zero temperature Drude weight should
consequently be finite in the
thermodynamic limit, yet the contribution of the Drude weight relative to the regular
conductivities remains an open issue for low temperatures.

In the case of heat transport [Fig.~\ref{fig:DrudeI0}(b)], the picture 
has more facets. For low temperatures ($T=0.25J$, solid black line)
$D_{\mathrm{th}}/I^0_{\mathrm{th}}$ decreases in the field-induced phase as
there is a huge increase in the regular part of the conductivity [compare
Fig.~\ref{fig:conduct}(c)]. This results in a minimum in $D_{\mathrm{th}}/I_{\mathrm{th}}$ at $h\approx J$, roughly where the 
field-dependent thermal Drude weight has its maximum. At $h\approx 2J$, the Drude weight is restored as
the dominant contribution at $D_{\mathrm{th}}/I^0_{\mathrm{th}} \approx 0.9$ at low
$T$. This effect is weakened by temperature [compare Fig.~\ref{fig:conduct}(d)],
eventually resulting in a monotonous increase of
$D_{\mathrm{th}}/I^0_{\mathrm{th}}$ with $h$ at $T=J$. Our results clearly
suggest that the Drude weight is a significant contribution in the field-induced
gapless phase on {\it finite} systems. Several comments are in order. First, a finite Drude weight at $T>0$ in a non-integrable system would be surprising, while on the other hand, this observation is
consistent with the fact that thermodynamic properties in the phase are well
described by an effective $XXZ$ model, \cite{totsuka98,klanjsek08} which is believed to have ballistic transport properties in its gapless zero-field phase.\cite{hm07a,zotos99} 
Moreover, in
the ferromagnetic phase $h>h_{c_2}$, the physics is expected to be well
approximated by a weakly-interacting gas of magnons which in turn renders the
Drude weight large in relation to the total weight. We stress that similar to $D_{\mathrm{s}}$, this is a statement about finite
systems and temperatures, and much larger system sizes (see Appendix) would be required to extrapolate the ratio
$D_{\mathrm{th}}/I^0_{\mathrm{th}}$ to the thermodynamic 
limit. We can therefore not pursue this question here. Note though, that in the
high-temperature limit ($\beta=0$) the magnetic field dependence drops out. In
this limit, exact diagonalization yields a systematic decrease of the Drude weight with the
system size.\cite{hm03}

To summarize the discussion of the behavior at $\lambda=0.5$, in the
low-temperature regime, which is our main case of interest, the field-driven
enhancement is visible in the conductivities as well as in the Drude weights and
therefore also in the spectral weight, but the Drude weights contribute the
most. Further, we can distinguish a field- and a temperature-dominated regime:
qualitatively, at $T<h_{c_1}$, a variation
of the magnetic field influences the conductivities, whereas for $T>h_{c_1}$,
the magnetic field has little effect on both the structure and the weight in the
conductivities. Figure~\ref{fig:I05} contains the main result of our work: a field-induced increase
in both transport coefficients at low frequencies $\omega <h_{c_1}$.

\subsection{Dependence on $\lambda$}
\label{sec:lambda}

\begin{figure*}[t]
\includegraphics[width=0.45\textwidth,angle=0]{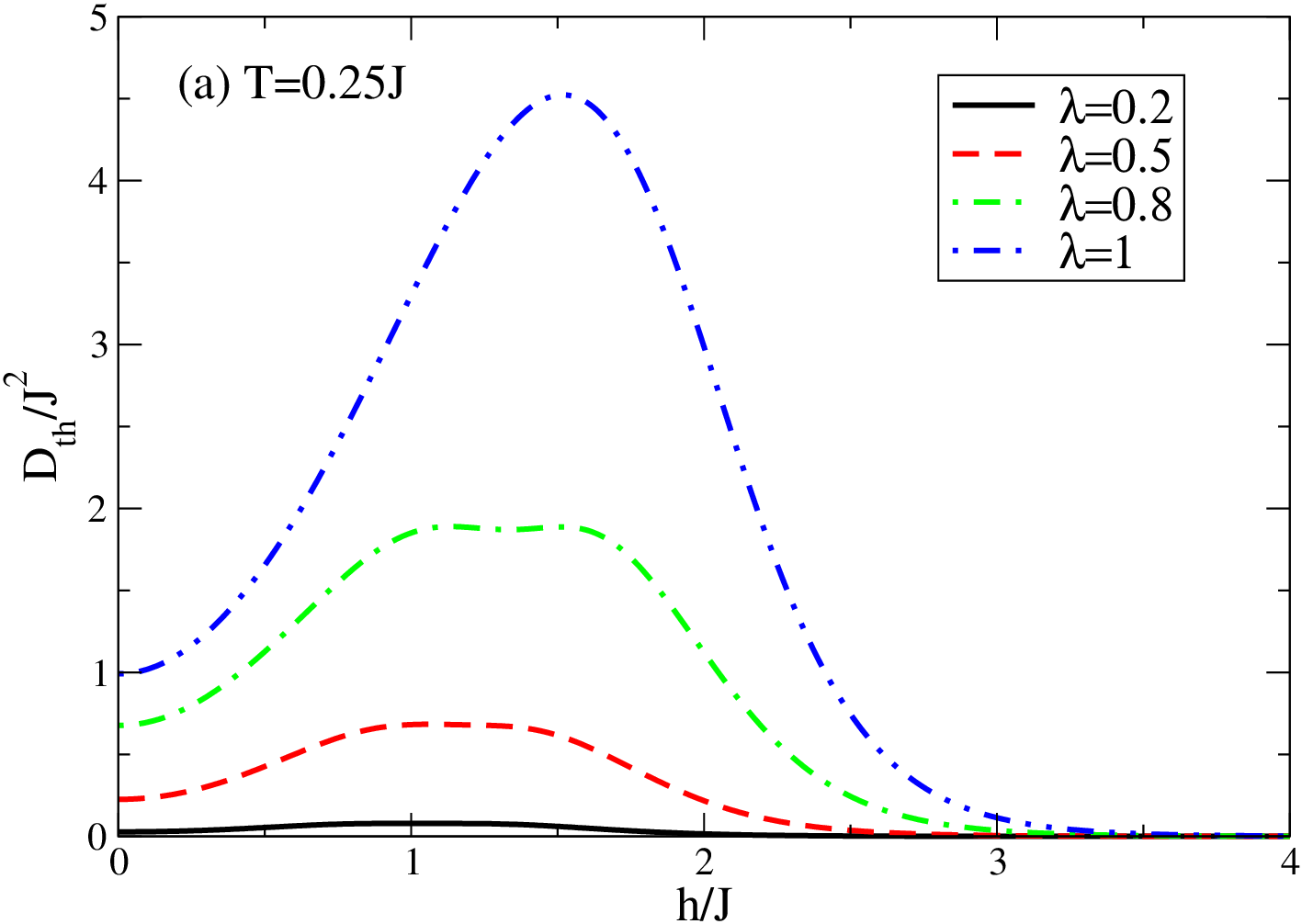}
\includegraphics[width=0.45\textwidth,angle=0]{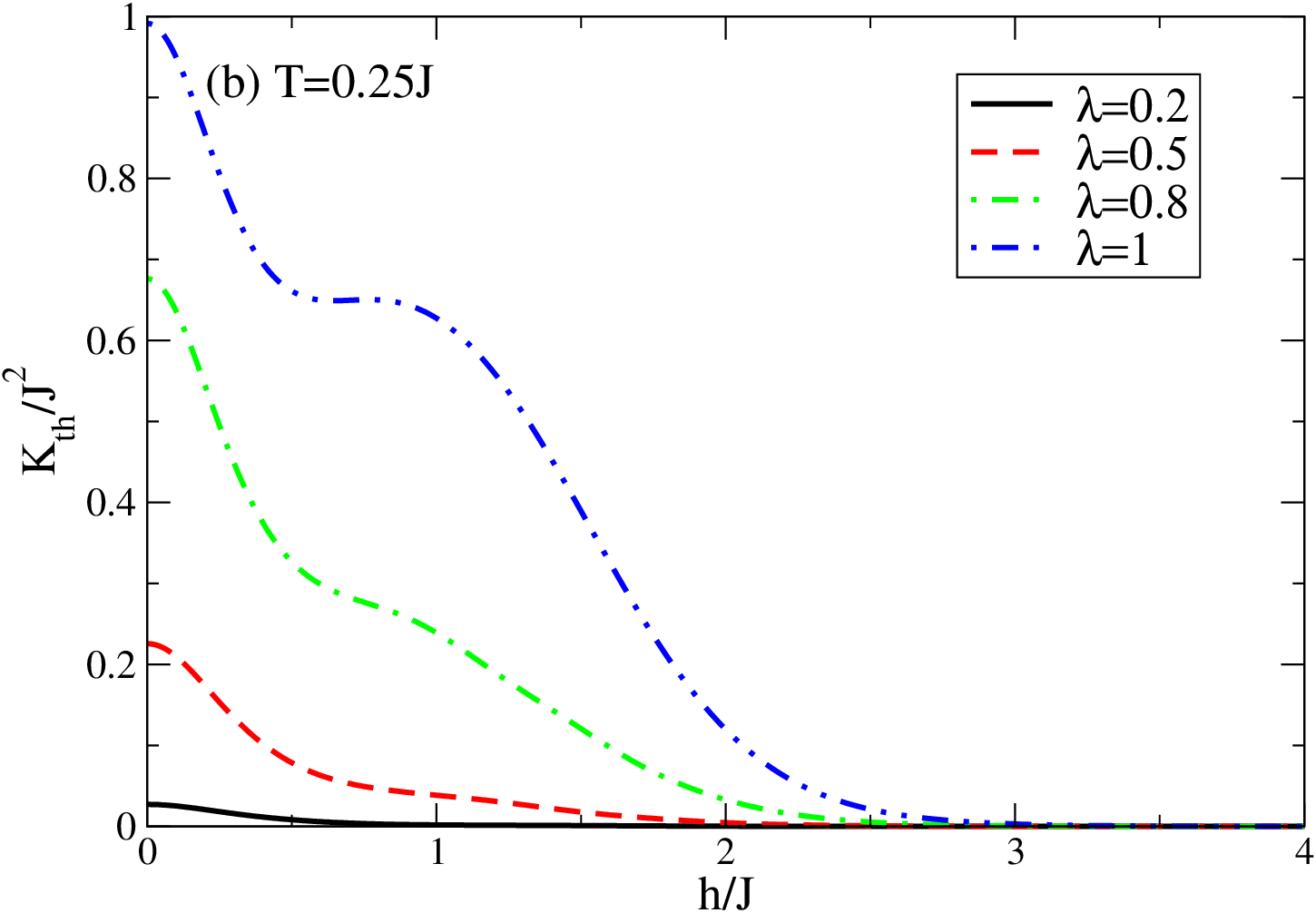}
\caption{(Color online) (a) Thermal Drude weight $D_{\mathrm{th}}$ as a function of magnetic field $h$ for a system of length $L=16$ and different values of $\lambda$ at low temperatures ($T=0.25J$). While the thermal Drude weight decreases fast as $\lambda$ decreases, the field dependence is not affected. (b) Corrected thermal Drude weight $K_{\mathrm{th}}$. We find that the magnetothermal coupling prohibits the strong increase of the thermal Drude weight in the field-induced phase.}
\label{fig:thermomag}
\end{figure*}

Next we turn to the discussion of the influence of the strength of dimerization
$\lambda$. Since we have presented evidence that the Drude weights are the
dominant contribution in the field-induced phase, we concentrate on these quantities,
expecting them to reflect the main qualitative behavior.
Figure~\ref{fig:drudelam} shows the spin Drude weight as a function of the
external magnetic field at different temperatures and strengths of dimerization.
In all cases, the Drude weight for spin transport exhibits a maximum at
intermediate fields. We expect that in the thermodynamic limit and at sufficiently low temperatures,
the transition into the field-induced Luttinger phase should lead to signatures
in the Drude weights at $h_{c_1}$ and $h_{c_2}$. Yet, for the system sizes we can study, and thus the accessible $T$,
the location of the inflection points of $D_{\mathrm{s}}$ do not exhibit a clear
correlation with critical fields $h_{c_1}$ and $h_{c_2}$ of Fig.~\ref{fig:phase}. This remains to be analyzed in
the future.

As temperature increases [compare Fig.~\ref{fig:drudelam}(a) and
Fig.~\ref{fig:drudelam}(b)], the maximum in the Drude weights moves from between
the critical fields to higher values of $h$ while the absolute value of
$D_{\mathrm{s}}$ decreases. Sending $\lambda\to 0$ decreases the values of
$D_{\mathrm{s}}$ for all fields, while the field-driven increase is not affected
qualitatively. This is true for small as well as high
temperatures. Note that increasing the temperature suppresses the spin
Drude weight much more severely than altering $\lambda$. In the case of thermal
transport [Fig.~\ref{fig:thermomag}(a)] we
find a stronger quantitative dependence on $\lambda$ while the fact that altering
$\lambda$ does not change the field dependence remains qualitatively 
correct. The former is not surprising, since uncoupled dimers cannot carry any heat
current while the Heisenberg chain is known for its ballistic heat
transport.\cite{zotos97,kluemper02,sakai03,hm03}

\subsection{Magnetothermal couplings}
\label{sec:thermomag}

\begin{figure}[t]
\includegraphics[width=0.45\textwidth,angle=0]{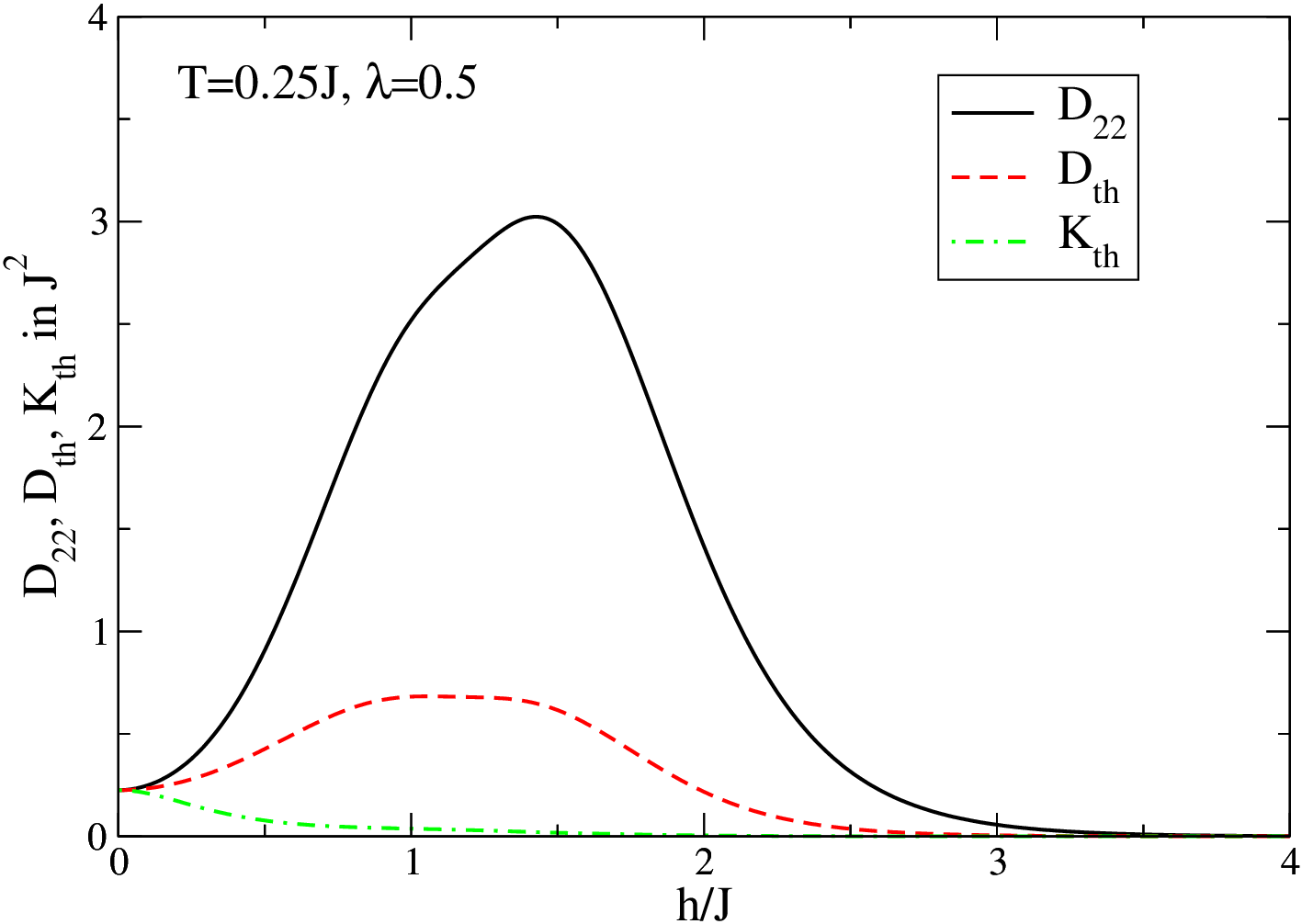}
\caption{(color online) Comparison of the various thermal Drude weights $D_{\mathrm{th}}$, $D_{22}$ [Eq.~\eqref{eq:w4} and [Eq.~\eqref{eq:D22}], $K_{\mathrm{th}}$ [Eq.~\eqref{eq:thmag1}] 
for thermal transport at $\lambda=0.5$ and $T=0.25J$ for a chain of length $L=16$. 
While the three curves coincide at $h=0$, the field dependence exhibits huge differences. On the one hand, $D_{22}$ given by Eq.~\eqref{eq:D22} (solid black line) shows a much stronger enhancement than $D_{\mathrm{th}}$ (dashed red line). On the other hand, including the magnetothermal correction (dot-dashed green line) leads to a monotonous decrease of $K_{\mathrm{th}} $ as the magnetic field increases.
}
\label{fig:D22}
\end{figure}

The last effect to investigate is the influence of the magnetothermal coupling Eq.~\eqref{eq:thmag1}. In the case of the $XXZ$ chain, 
it has been shown that this correction suppresses the thermal conductivity.\cite{hm05} The results for the field-dependence of the thermal Drude weight $K_{\mathrm{th}}$ at low temperatures ($T=0.25J$) are shown in Fig.~\ref{fig:thermomag}(b). The corrected Drude weight $K_{\mathrm{th}}$ [Eq.~\eqref{eq:thmag1}] decreases as the magnetic field increases, which is the systematic behavior
in the whole field-induced phase. 
Shifting $\lambda$ towards $\lambda=0 $ lowers the overall values while only slightly affecting the field dependence. 
Figure~\ref{fig:D22} is a comparison of the three Drude weights associated with thermal transport, $D_{\mathrm{th}}, D_{22}$ and $K_{\mathrm{th}}$, where
\begin{equation}
D_{22}= D_{\mathrm{th}}-2\beta hD_{\mathrm{th,s}}+\beta h^2 D_{\mathrm{s}} \label{eq:D22}\,.
\end{equation}
 As expected from Eqs.~\eqref{eq:thmag1} and \eqref{eq:D22}, all three quantities coincide at zero magnetic field. While $D_{22}>D_{\mathrm{th}}$ both quantities exhibit a similar field dependence , i.e., it features a maximum in the field-induced gapless phase and a smooth decay to zero as $h$ increases beyond saturation. The magnetothermal correction changes this behavior completely, yielding a monotonous decrease as the field increases. 

\section{Conclusions and summary}
\label{sec:conc}

In this work we studied spin and heat transport in dimerized spin$-1/2$ chains
in a magnetic field, in dependence of temperature $T$, magnetic field $h$ and
the strength of dimerization $\lambda$. Focusing on the field dependence in the
case of $\lambda=0.5$ we found that the transport coefficients at low
frequencies for both, spin and heat transport, are strongly enhanced in the
field-induced gapless phase. For spin transport at low but finite temperatures the Drude
weight becomes the
dominant contribution as the field increases. 
We stress that this an observation for finite systems. While one may expect that the Drude weight remains
relevant in the field-induced phase in the zero temperature,\cite{kohn64} thermodynamic
limit, the theoretically interesting question of a finite Drude weight at low but finite temperatures
is beyond the scope of this work.

In the case of heat transport, the emerging picture is more involved. 
 Increasing the magnetic field up to saturation, we find that the
regular part of the conductivity is vastly enhanced at all frequencies. Although
the thermal Drude weight has a field-dependence similar to its spin counterpart,
this leads to a decrease of the relative contribution in the spectral weight at
low temperatures. However, the thermal Drude weight remains the dominant
contribution on the finite systems studied here. In both transport channels, a temperature larger than the spin
gap severely weakens all features related to the magnetic field.

\begin{figure*}[tb]
\includegraphics[width=0.45\textwidth,angle=0]{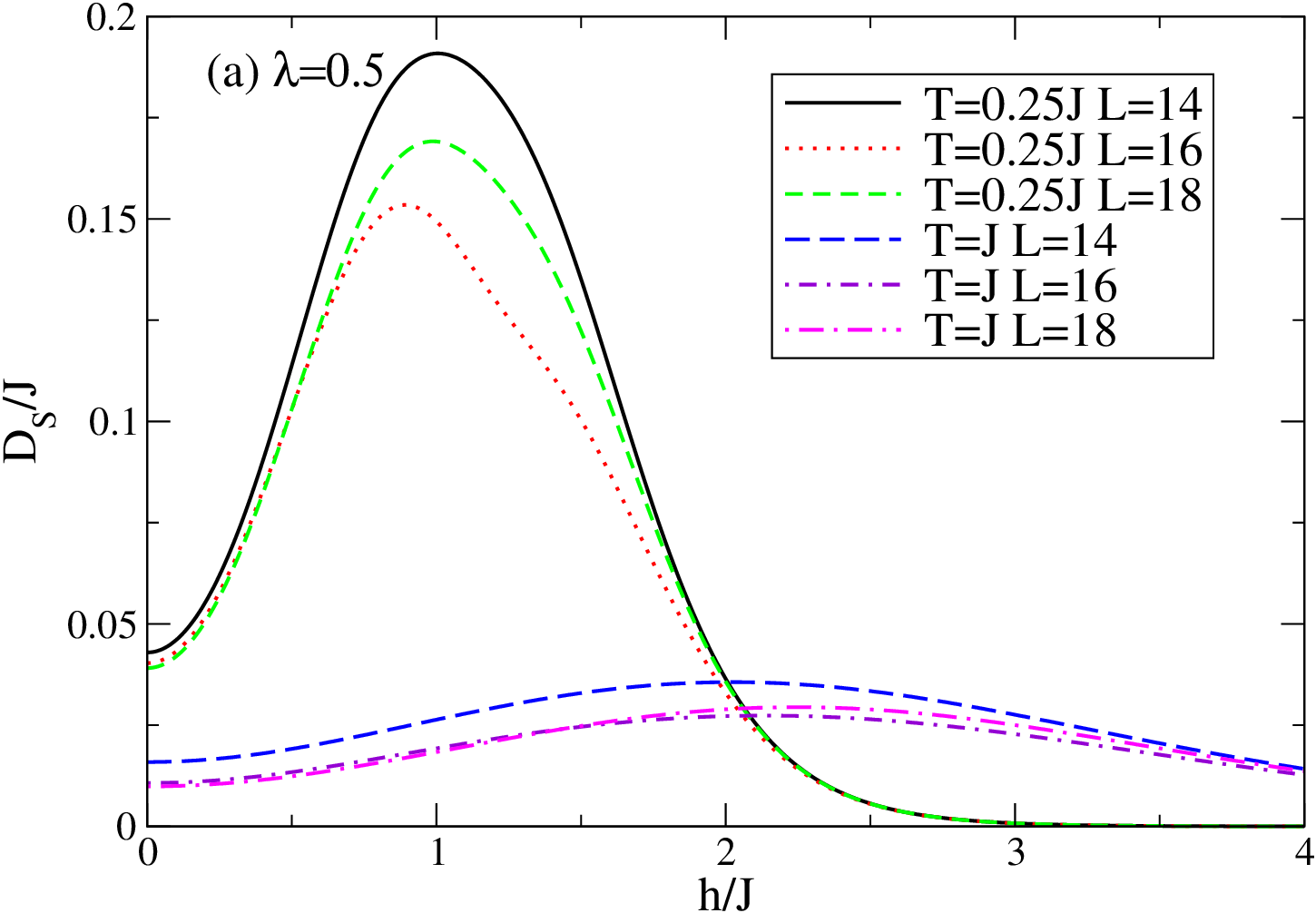}
\includegraphics[width=0.45\textwidth,angle=0]{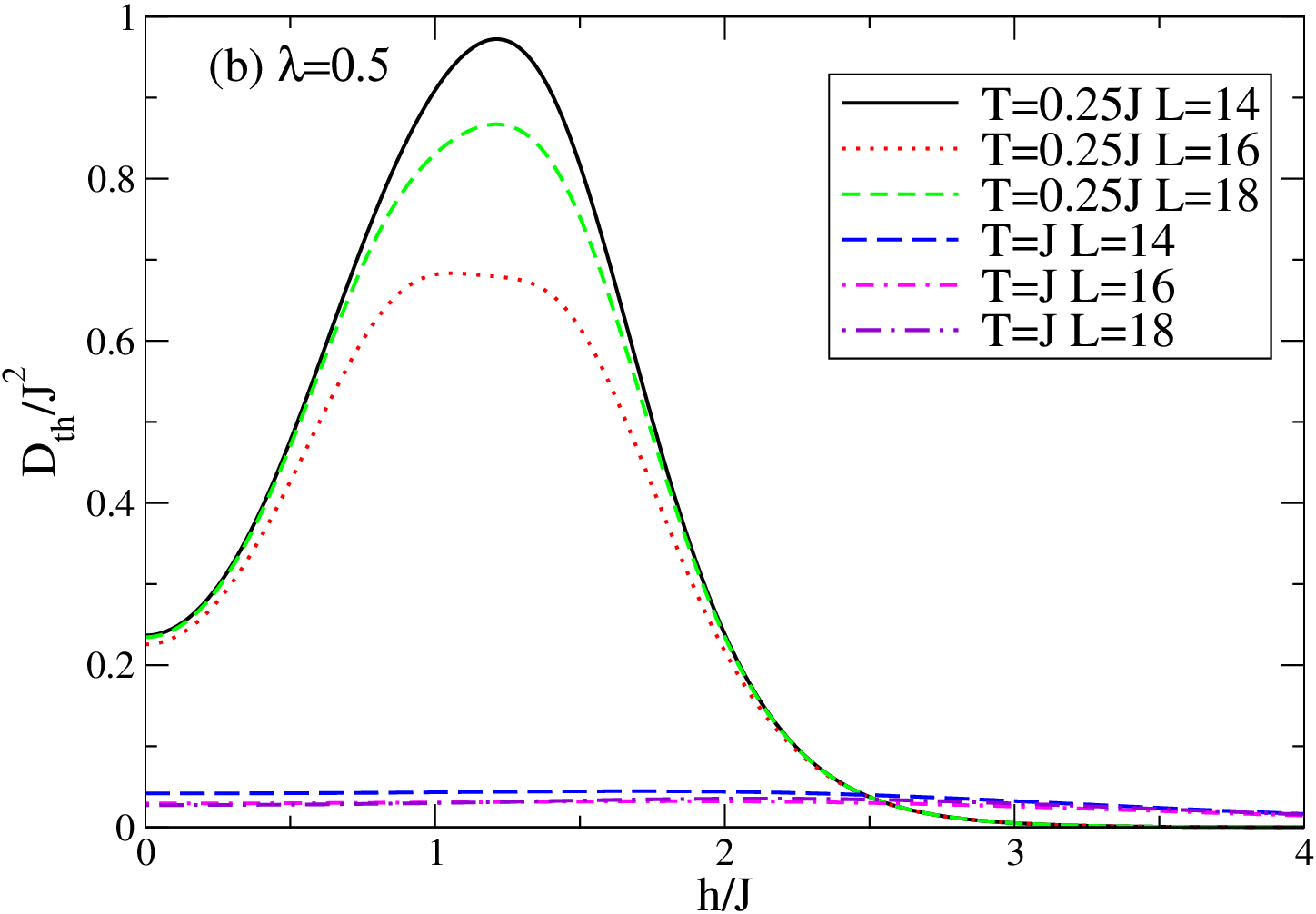}
\caption{(color online) Finite-size effects in the field-dependent spin Drude weight (a) and thermal Drude weight (b) at two different temperatures ($T=0.25J,J$) for the system sizes $L=14,16,18$.}
\label{fig:App1}
\end{figure*}

Our main finding is the field-induced increase of both transport coefficients in the
low-frequency window, which is robust even on the finite chains accessible to exact
diagonalization. While most of our work was concerned with parameters that mimic
the energy scales typical for (C$_{5}$H$_{12}$N)$_{2}$CuBr$_{4}$ [Ref. \onlinecite{thielemann09,ruegg08,klanjsek08}], 
we further studied the dependence on the strength of dimerization. There we
observe that altering $\lambda$ changes the overall values while the field
dependence remains qualitatively the same. Thus we conclude that our observation
of field-enhanced spin and heat transport is valid in dimerized spin
chains. We also expect it to be more generally valid in other dimerized quasi one-dimensional spin systems, such as two-leg spin ladders.

Finally, the calculation of the magnetothermal coupling at low temperatures
unveils a very interesting effect, similar to the $XXZ$ chain.\cite{hm05} The corrected thermal Drude weight shows no
increase in the field induced phase, instead, it decreases.

In conclusion, we complemented several experimental and theoretical studies which characterized the field-induced gapless phase,
\cite{thielemann09,ruegg08,klanjsek08} by emphasizing here that clear fingerprints of this transition are present in current-current correlation functions and should thus, in principle, manifest themselves in transport measurements.The spin conductivity is inaccessible at the moment, but can be extracted from quantities measured in NMR experiments.\cite{takigawa96,thurber01}
 
\begin{acknowledgements}
We are grateful to Andreas Honecker and Robin Steinigeweg for fruitful discussions. This work was supported by the { \it Deutsche Forschungsgemeinschaft} through FOR 912.
\end{acknowledgements}

\appendix

\section{Finite size scaling of the Drude weights}

The finite-size scaling of the Drude weights in non-integrable spin chains has been previously studied at zero field and in the limit of infinite temperature \cite{hm03} without evidence of a finite Drude weight in the thermodynamic limit for non-integrable systems. Note that the field dependence drops out at $\beta=0$. These studies include the dimerized chain. We present our data for finite magnetic fields in Fig.~\ref{fig:App1}. Due to the interplay between the regular contribution and the Drude weight in the conductivities, the dependence on the system size is non-monotonic at finite magnetic fields and for the system sizes accessible, especially at low temperatures $(T=0.25J)$. Therefore, the accessible system sizes are too small to gain any qualitative insight into $D_{\mathrm{s[th}]}$ beyond Ref.~\onlinecite{hm03}. The integrated spectral weight (Fig. \ref{fig:I05}) is a robust quantity in this context.


\end{document}